\documentclass[journal]{new-aiaa}
\usepackage[utf8]{inputenc}
\usepackage{textcomp}
\usepackage[permil]{overpic}
\usepackage{graphicx}
\usepackage{amsmath}
\usepackage[version=4]{mhchem}
\usepackage{siunitx}
\usepackage{longtable,tabularx}
\usepackage{float}
\usepackage{xcolor}					

\title{Duty-cycle actuation for drag reduction of deep dynamic stall: \\ Insights from linear stability analysis}

\author{Lucas Feitosa de Souza \footnote{Graduate Student, School of Mechanical Engineering, University of Campinas, lucasfeitosa010@gmail.com} and William Wolf.\footnote{Associate professor, School of Mechanical engineering, University of Campinas, wolf@fem.unicamp.br}}
\affil{University of Campinas, Campinas, SP, 13083-860, Brazil}
\author{Maryam Safari\footnote{Graduate Student, Department of Mechanical and Aerospace Engineering, North Carolina State University, msafari2@ncsu.edu } and Chi-An Yeh \footnote{Assistant Professor, Department of Mechanical and Aerospace Engineering, North Carolina State University, chian.yeh@ncsu.edu}}
\affil{North Carolina State University, Raleigh, NC, 27695, USA}

\begin{document}

\maketitle

\begin{abstract}
A flow control framework based on linear stability analysis is proposed focusing on reducing 
the aerodynamic drag due to dynamic stall through a finite-window temporal actuation. 
The methodology is applied on a periodically plunging SD7003 airfoil.
Finite-time Lyapunov exponent (FTLE) fields reveal a saddle point near the 
airfoil leading edge, where a shear layer forms and feeds a dynamic stall vortex (DSV). 
A local stability analysis conducted at this saddle point identifies a Kelvin-Helmholtz 
instability, and the most unstable eigenvalue frequencies remain constant when the variation
in the effective angle of attack is minimal. The findings from the FTLE fields and the stability
analysis are used to inform the position and finite duty cycle of a periodic blowing and suction actuation
applied in a wall-resolved large eddy simulation (LES). The present framework reduces the actuation duty cycle by 77.5\% during the airfoil plunging motion, while maintaining the same performance as a continuous 
actuation throughout the entire cycle. The LES results demonstrate that disturbances from the 
stability-analysis-informed actuation modify the leading-edge dynamics, preventing the formation
of the coherent DSV and significantly reducing the drag.
\end{abstract}



\section{Introduction}

Dynamic stall impacts a wide range of engineering systems such as helicopters in forward flight, drones, wind turbines, and aircraft undergoing rapid maneuver. It is characterized by an unsteady boundary layer separation and the emergence of a leading edge vortex, also called the dynamic stall vortex (DSV). 
During its formation and transport, the DSV induces high unsteady loads that reduce the airfoil aerodynamic performance, potentially compromising the structural integrity of wings and blades \cite{corke2015dynamic}.

Flow control is a critical topic in dynamic stall research and it aims to prevent or mitigate the effects of the DSV. Advances in numerical simulations and image-based experimental techniques enabled the acquisition of velocity fields with high temporal and spatial resolution. Such capabilities allowed the study of viscous flow phenomena near the airfoil surface, deepening our understanding of boundary layer separation mechanisms \cite{Mulleners2012, gupta_ansell_2019, Benton_Visbal_2019, Ansell_Karen_2020, Ansell_2020_evolution, Miotto_Wolf_Gaitonde_Visbal_2022}. This, in turn, permitted novel flow control strategies aiming to prevent the formation of the DSV, or reduce its impact on the aerodynamic loads \cite{visbal_ctrl,Visbal_benton_2018, Ramos_PRF, Visbal2023_cavity, GARDNER2023100887,LeFouest2024}.

At moderate Reynolds numbers, the onset of dynamic stall is attributed to bursting of a laminar separation bubble that forms in the airfoil leading edge \cite{Benton_Visbal_2019,  gupta_ansell_2019}. Through knowledge of the bubble characteristic frequencies, control strategies have been proposed to suppress the bursting phenomenon and keep the flow attached along the entire airfoil motion. The success of this strategy was shown by considering active control through blowing and suction \cite{Visbal_benton_2018,visbal_ctrl}, and passive control with properly sized cavities placed at the airfoil leading edge \cite{Visbal2023_cavity}. The phenomenon of dynamic stall has also been investigated for cases where other transition mechanisms are relevant. In the absence of a laminar separation bubble, investigations point to Kelvin-Helmholtz instabilities arising at the leading edge as a mechanism for separation of the boundary layer and formation of the DSV \cite {Visbal_2011, Miotto_Wolf_Gaitonde_Visbal_2022}. For such cases,  \citeauthor{Ramos_PRF} \cite{Ramos_PRF} showed that low frequency blowing and suction can be effective to disrupt the DSV formation and mitigate the drag caused by its advection. 

With the purpose of testing flow actuation approaches, several studies focused on identifying coherent structures related to the dynamic stall onset. In this context, flow modal decomposition techniques such as proper orthogonal decomposition (POD), dynamic mode decomposition (DMD), empirical mode decomposition (EMD) and Lagrangian coherent structures (LCS) stand out in the literature \cite{Ansell_Karen_2020, Ansell_2020_evolution, Mulleners2012, deSouza2024, Miotto_Wolf_Gaitonde_Visbal_2022, Dunne2015}. However, for convection dominated flows with intermittent features such as dynamic stall, it can be difficult to interpret the spatial support of modes generated from techniques relying on the singular value decomposition \cite{Miotto_Wolf_Gaitonde_Visbal_2022,deSouza2024}. In these cases, the causal relationship between the observed modes and control success may be compromised.

Techniques based in a deeper physical foundation, such as stability and resolvent analyses, are also applied aiming to provide flow control strategies for airfoils under static and dynamic stall \cite{Yeh_Taira_2019, Chi_an_2020, Kern_Negi_Hanifi_Henningson_2024}. These methods provide the dynamics of linear disturbances around a base flow that satisfies the Navier-Stokes equations. The success of this type of analysis in producing valuable insights of the flow physics, transition mechanisms, besides flow actuation strategies has already been proven for cases with a time-invariant base flow \cite{Ricciardi_Wolf_Taira_2022, Ribeiro_Yeh_Taira_2023, Chi_an_2020,  Yeh_Taira_2019}. Recently, \citeauthor{Ribeiro_Yeh_Taira_2023} \cite{Ribeiro_Yeh_Taira_2023} employed triglobal resolvent analysis to investigate the effects of wingtip and sweep angle on the wake of swept wings.  \citeauthor{Chi_an_2020} \cite{Chi_an_2020} used resolvent analysis to design active control strategies for the separated flow over a NACA0012 airfoil, reducing drag by up to 49\% and increasing lift by up to 54\%.

Considering time-variant systems, different approaches can be found in the literature to extend the application of linear analysis techniques. \citeauthor{Padovan_Otto_Rowley_2020} \cite{Padovan_Otto_Rowley_2020} extended the resolvent analysis to time periodic base flows aiming to study interactions between different frequencies in the wake of an airfoil. \citeauthor{Kern_Negi_Hanifi_Henningson_2024} \cite{Kern_Negi_Hanifi_Henningson_2024} used the optimally time dependent (OTD) framework from \citeauthor{Babaee_Sapsis} \cite{Babaee_Sapsis} to investigate the temporal evolution of the laminar separation bubble that forms on the leading edge of an airfoil for low-amplitude pitching motions, revealing its unstable character. \citeauthor{Ansell_2020_evolution} \cite{Ansell_2020_evolution} applied linear stability analysis through the solution of the Orr-Sommerfeld equation to study the mechanism that gives birth to coherent structures in the shear layer forming on the leading edge of a NACA0012 under dynamic stall. Using phase-averaged velocity profiles obtained by a time-resolved particle image velocimetry field, it was shown that, for small variations in the angle of attack, the scatter of the eigenvalue spectra due to the time-dependent base flow is small. By analysing the most unstable eigenvalues, they were able to identify frequencies related to the roll-up of Kelvin-Helmholtz instabilities described in the literature for canonical shear layer profiles.

Despite recent efforts, most of the work applying linear techniques to dynamic stall is limited to characterization of the flow unsteady behavior. In the present work, we combine linear stability analysis and active flow control for deep dynamic stall. A blowing and suction jet is applied at the airfoil leading edge to disrupt the formation of the DSV, which is fed by a shear layer. The choice of actuation placement is justified by the identification of potential saddle points of the flow system through computation of finite-time Lyapunov exponents (FTLEs). A local stability analysis is employed to identify the most unstable frequencies of the leading edge shear layer. Moreover, it also determines the timing to turn on and off the actuation in order to reduce the DSV-induced drag using a finite-window duty cycle for the actuator. Results show that even a simplified local analysis can provide important insights to design a control strategy that improves aerodynamic performance during dynamic stall.

\section{Numerical Methods}\label{sec2}

\subsection{Large eddy simulations}

High-fidelity numerical simulations are performed for an SD7003 airfoil undergoing five plunging cycles. The vertical displacement of the airfoil is defined as $ h(t) = h_0 \sin{(\kappa t)}$, where $t$ is the non-dimensional time, $\kappa$ is the reduced frequency, and $h_0$ is the plunge amplitude. The flow has a freestream Mach number $M_\infty = 0.1$ and a Reynolds number of $Re = 6 \times 10^4$. All parameters are nondimensionalized by the freestream velocity $U_\infty$ and airfoil chord $c$. Similarly to Refs. \cite{Visbal_2011, Ramos_PRF}, the motion parameters are set as $\kappa = 0.5$ and $h_0 = 0.5$. In the present plunging airfoil, the effective angle of attack depends on the static geometrical angle ($AoA_{\text{st}}$) besides the apparent incoming flow induced by the vertical motion and freestream velocity as

\begin{equation}
    AoA(t) =AoA_{\text{st}} -\text{tan}^{-1}
    \left[\frac{\dot{h}(t)}{U_\infty}\right] \mbox{ .}
\end{equation}
Considering the motion parameters employed and  $AoA_{\text{st}} = 8^{\circ}$, the airfoil experiences a variation in effective angle of attack within the range $-6^{\circ} \leq AoA \leq 22^{\circ}$ during a plunging cycle of $4\pi$ ($\approx 12.56$) convective time units.

For the simulations, an O-type grid is employed and the equations are solved in a general curvilinear coordinate system. The same grid used in Ref. \cite{Ramos_PRF} is employed in this work for all simulations, having $480 \times 350 \times 96$ points in the streamwise, wall-normal, and spanwise directions, respectively. This grid was shown in the previous reference to produce converged statistics validated against the literature \cite{Visbal_2011,Miotto_Wolf_Gaitonde_Visbal_2022}. In order to resolve the most energetic flow scales and capture viscous phenomena near the airfoil surface, wall-resolved LES are conducted. The spatial discretization of the flow equations is performed using the sixth-order accurate compact finite-difference scheme from Nagarajan et al. \cite{Nagarajan2003}. This methodology solves the advective and viscous fluxes using a staggered grid setup. The sixth-order compact interpolation scheme presented in the previous reference is also employed to interpolate flow quantities between collocated and staggered grids.

The time integration of the flow equations is performed using an explicit third-order compact-storage Runge-Kutta scheme in regions away from solid boundaries. Near the airfoil surface, an implicit second-order scheme is applied to overcome the stiffness problem typical of boundary layer grids. Sponge layers and characteristic boundary conditions based on Riemann invariants are applied in the farfield, and adiabatic no-slip boundary conditions are used along the airfoil surface. At the leading edge we employ a blowing and suction actuator with a jet speed corresponding to 20\% of the freestream velocity. The actuation moment coefficient is given by $C_\mu = 1.12 \times 10^{-2} \%$, and this setup corresponds to case 3 (weakest actuation) employed by \citeauthor{Ramos_PRF} \cite{Ramos_PRF}. More details about the actuation function will be given in the following sections. The present numerical tool has been validated against experimental results and high-fidelity numerical simulations of turbulent flows \cite{Nagarajan2003, Bhaskaran, wolf2012}, besides the study of dynamic stall \cite{Ramos_PRF, Miotto_Wolf_Gaitonde_Visbal_2022, Miotto_pitch_plunge}.

\subsection{Linear stability analysis}

A local linear stability analysis is carried out considering the incompressible Navier-Stokes equations. In the linear analysis, the physical quantities are also nondimensionalized by freestream velocity and airfoil chord in order to compute frequencies comparable to those reported by  \citeauthor{Ramos_PRF} \cite{Ramos_PRF}. The wall normal derivatives $\mathbf{D} = \partial / \partial y$ and  $\mathbf{D^{2}} = \partial^{2} / \partial y^2$ are discretized using a sixth-order finite difference scheme. Dirichlet boundary conditions are imposed for all velocity disturbances. For pressure, we impose $\mathbf{D} \hat{p} = (1/\mbox{Re}) \mathbf{D^2} \hat{v}$ at the wall due to the no slip condition. The robustness of this choice of boundary condition is shown by \citeauthor{Hamada_Wolf_Pitz_Alves_2023} \cite{Hamada_Wolf_Pitz_Alves_2023}.

The flow disturbances are assumed to be periodic in the wall-tangential $x$ and spanwise $z$ directions, allowing the application of a Fourier transform as
\begin{equation}
    \hat{q}(x,y,z,t) =  q(y,t)\exp i (\alpha x + \beta z), \,\, q = (u, v, w, p)^T \mbox{ .}
\label{eq : fourier_space}
\end{equation}
In equation \ref{eq : fourier_space}, the symbols $\alpha$ and $\beta$ refer to the streamwise and spanwise wavenumbers, respectively, and $i$ is the imaginary number. The terms $u$, $v$, and $w$ represent the wall-tangential, wall-normal, and spanwise components of the velocity vector, and $p$ is the pressure. The coordinate system is shown in Fig. \ref{fig : profiles 1}. The process of linearization and non-dimensionalization generates the following system of equations for perturbations $q$ developing under a base flow $q_0 = (U(y), 0, 0, 0)$:

\begin{equation}
\begin{cases}
     \frac{\partial u}{\partial t} = \mathbf{A} \hat{u} - \mathbf{D} \mbox{U}\hat{v} - i \alpha \hat{p} \\
    \frac{\partial v}{\partial t} =  \mathbf{A} \hat{v} - \mathbf{D} \hat{p} \\
    \frac{\partial w}{\partial t} =  \mathbf{A} \hat{w} - i \beta \hat{p} \\
     0 = i \alpha \hat{u} + \mathbf{D} \hat{v} + i \beta \hat{w} \,\,\,\mbox{ ,} \\
\end{cases}
\label{eq : system}
\end{equation}
where the linear operator $\mathbf{A}$ is given by
\begin{equation}
    \mathbf{A} = -i \alpha \mathbf{U} + \frac{1}{Re} \left ( i^2 \alpha^2 + i^2 \beta^2  + \mathbf{D^2} \right) \mbox{ .}
\label{eq : LNS op}
\end{equation}

A Laplace transform is applied to equation \ref{eq : system} as
\begin{equation}
    \hat{q}(y,t) = \tilde{q}(y) \exp ( -i\omega t) \mbox{ ,}
\label{eq : laplace}
\end{equation}
allowing the system to be written in the following matrix form:
\begin{equation}
    -i \omega \mathbf{M} \tilde{q} = \mathbf{L} \tilde{q} \mbox{ .}
    \label{eq : eig}
\end{equation}
Here, $\omega$ is a complex number which provides the frequency and growth rate of the perturbations in the real and imaginary parts, respectively. 

The system given by equation \ref{eq : eig} can be solved as a generalized eigenvalue problem \cite{Hamada_Wolf_Pitz_Alves_2023}, where $\mathbf{M}$ and $\mathbf{L}$ are written as:
\begin{equation}
\mathbf{M} = 
\begin{vmatrix}
\mathbf{I} & 0 & 0 & 0 \\
0 & \mathbf{I} & 0 & 0\\
0 & 0 & \mathbf{I} & 0  \\
0 & 0 &   0 & 0     
\end{vmatrix} \mbox{ ,   }
    \mathbf{L} = 
\begin{vmatrix}
\mathbf{A} & -\mathbf{DU} & 0 & -i \alpha\\
0 & \mathbf{A} & 0 & \mathbf{-D}\\
0 & 0 & \mathbf{A} & -i \beta \\
i \alpha & \mathbf{D} & i \beta & 0
\end{vmatrix} \mbox{ .}
\end{equation}
Here, $\mathbf{M}$ is a singular matrix with the appropriate boundary conditions, and $\mathbf{I}$ is the identity matrix. In the present analysis, the Strouhal number is computed from the eigenvalues as:
\begin{equation}
     \lambda = eig(\mathbf{M},\mathbf{L}) = \omega_r + i\omega_i \implies  St = \frac{\omega_r c}{2 \pi U_{\infty}} \mbox{ ,}
\end{equation}
where the phase speed is $a_r$ and the growth rate is $a_i$ \cite{Schmid2012-wf}:
\begin{equation}
    a = \frac{\lambda}{\alpha} = a_r + i a_i \mbox{ .}
\end{equation}
The present analysis is conducted by identifying the Strouhal number and growth rate for the most unstable eigenvalue $\mbox{max}(a_i (t))$ at each time during the airfoil motion. 

\section{Results}\label{sec3}

\subsection{Flow analysis}

In this section, the flow over an SD7003 airfoil is analyzed. In order to describe the flow dynamics, contours of backward finite-time Lyapunov exponents (FTLEs) are computed for the baseline (uncontrolled) case. While the backward (negative) FTLE indicates regions where fluid particles are attracted in forward time, the forward (positive) FTLE depicts those regions where the particles are repelled the most \cite{Haller_annual_review, HALLER2000352,SHADDEN2005271, Green_2017, Green2010}. The computation of these quantities are achieved by integrating particle trajectories in forward and backward time to obtain the positive and negative FTLEs, respectively \cite{Brunton_LCS,SHADDEN2005271}.
The method employed for computing such quantities is discussed in the Appendix \ref{app:A}.

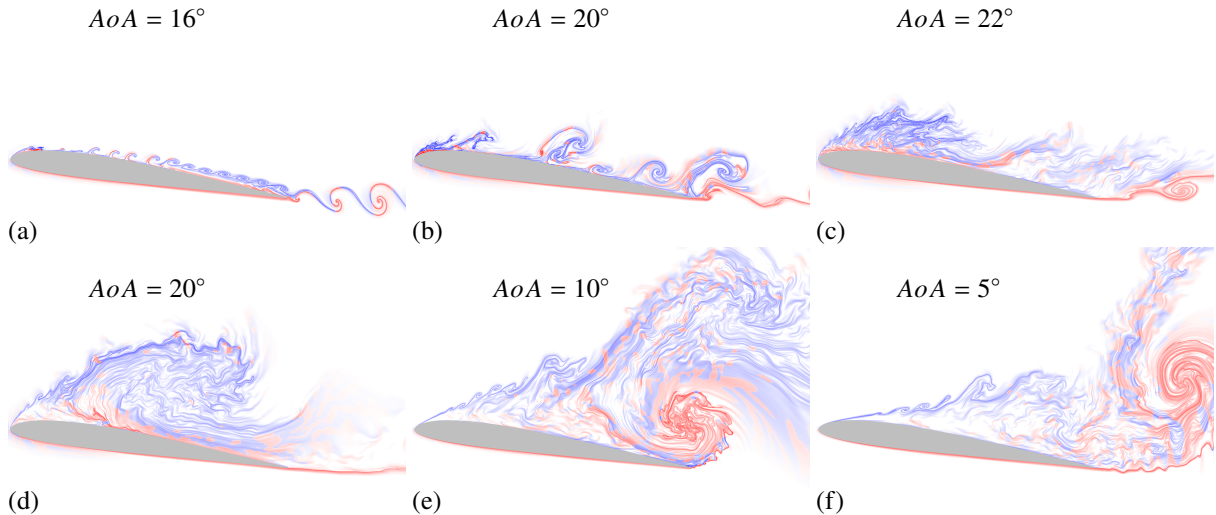
\begin{figure}[H]
     \centering
     \begin{overpic}[trim = 4.7cm 1cm 4cm 3cm,clip,width=0.32\textwidth]{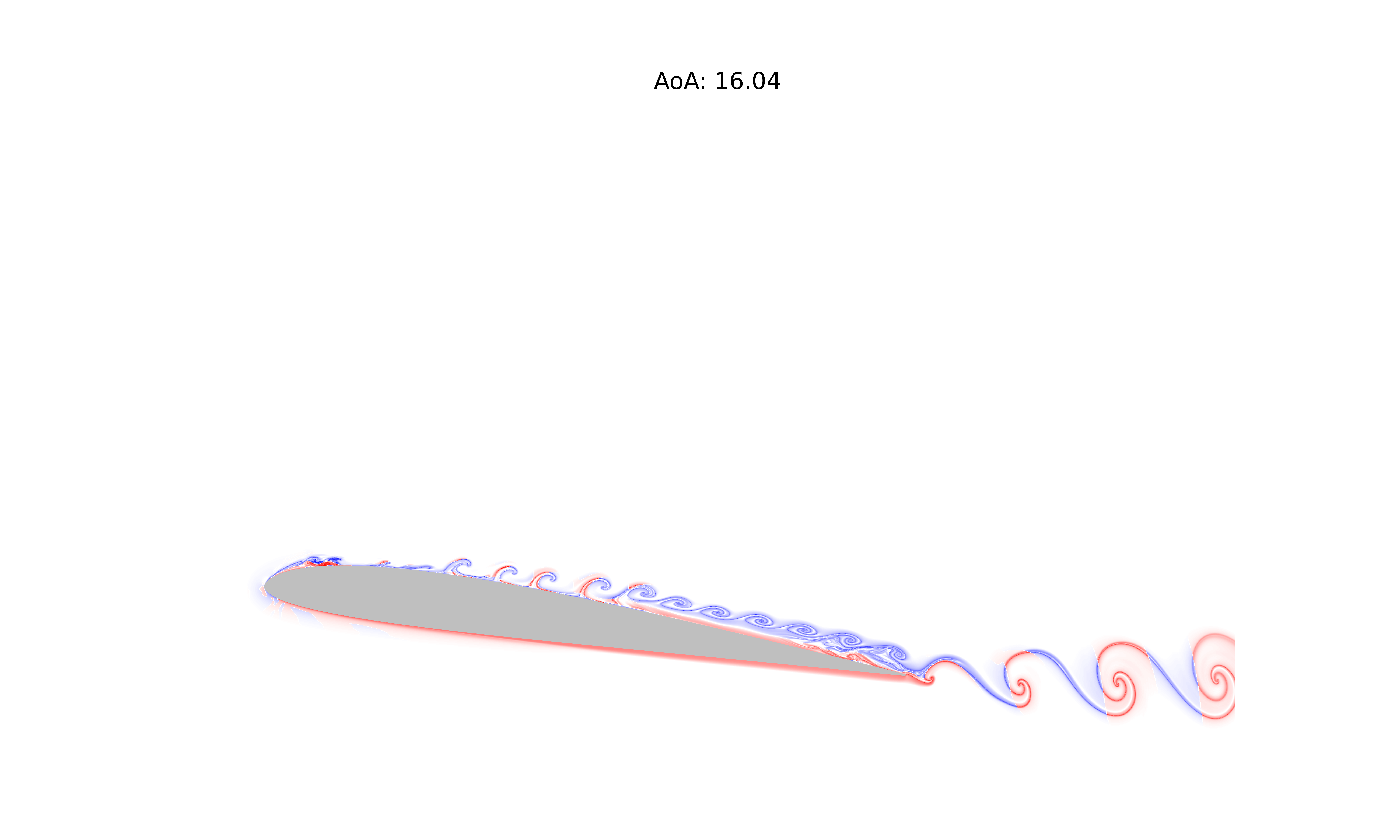}
        \put(0,15){(a)}
        \put(200,550){$AoA = 16^{\circ}$}
    \end{overpic}
    \begin{overpic}[trim = 4.7cm 1cm 4cm 3cm,clip,width=0.32\textwidth]{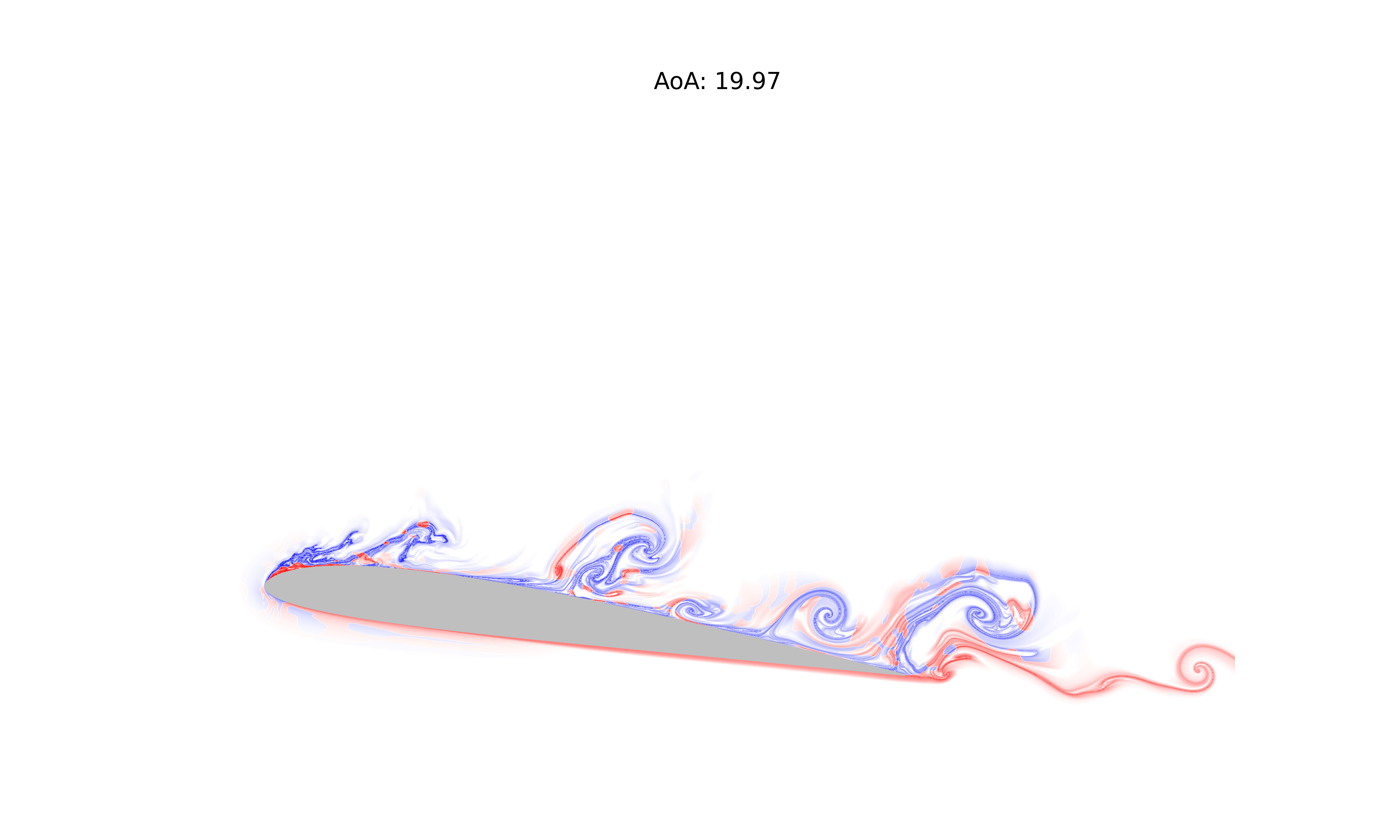}
        \put(0,15){(b)}
        \put(200,550){$AoA = 20^{\circ}$} 
    \end{overpic}
    \begin{overpic}[trim = 4.7cm 1cm 4cm 3cm,clip,width=0.32\textwidth]{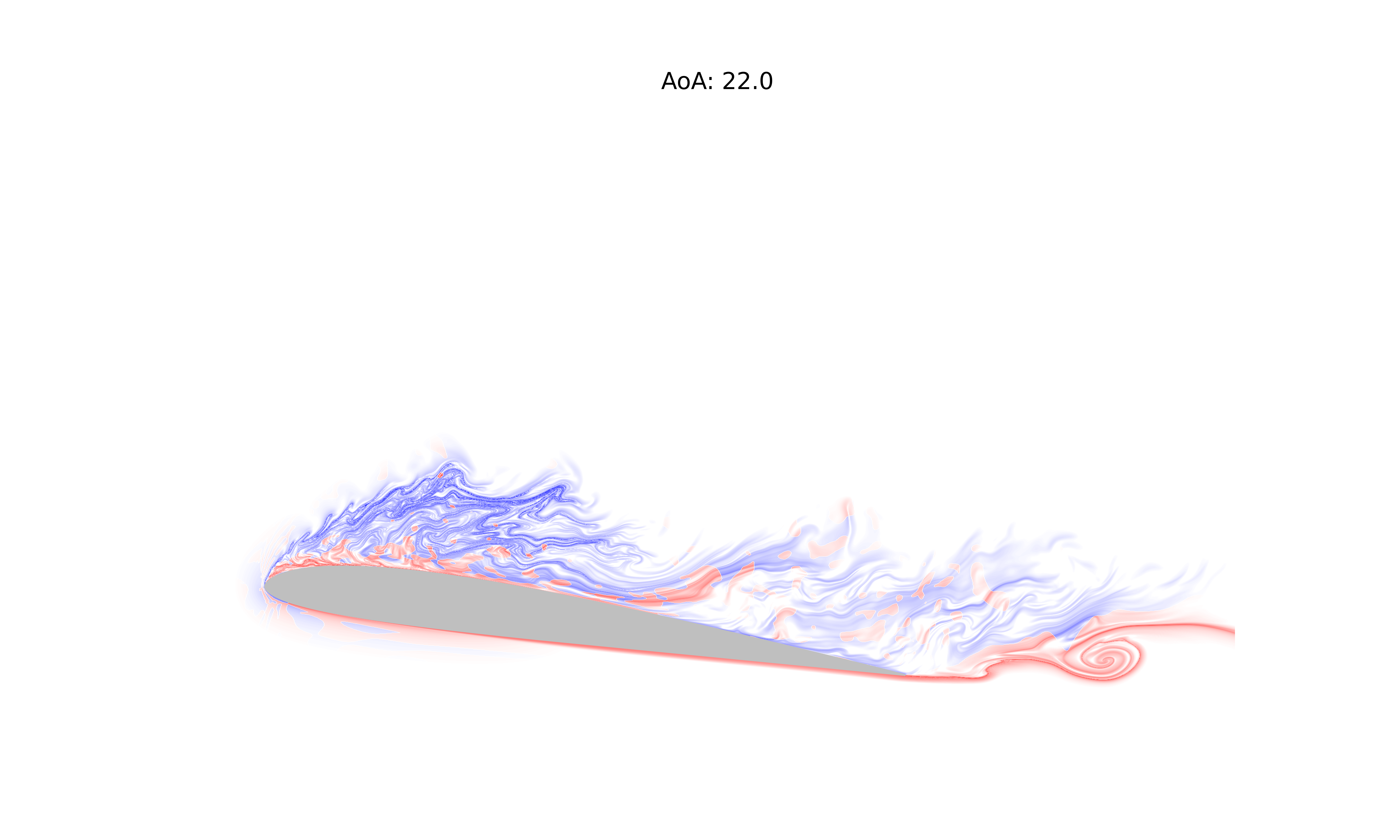}
        \put(0,15){(c)}
        \put(200,550){$AoA = 22^{\circ}$} 
    \end{overpic}
    \centering
     \begin{overpic}[trim = 4.7cm 1cm 4cm 3cm,clip,width=0.32\textwidth]{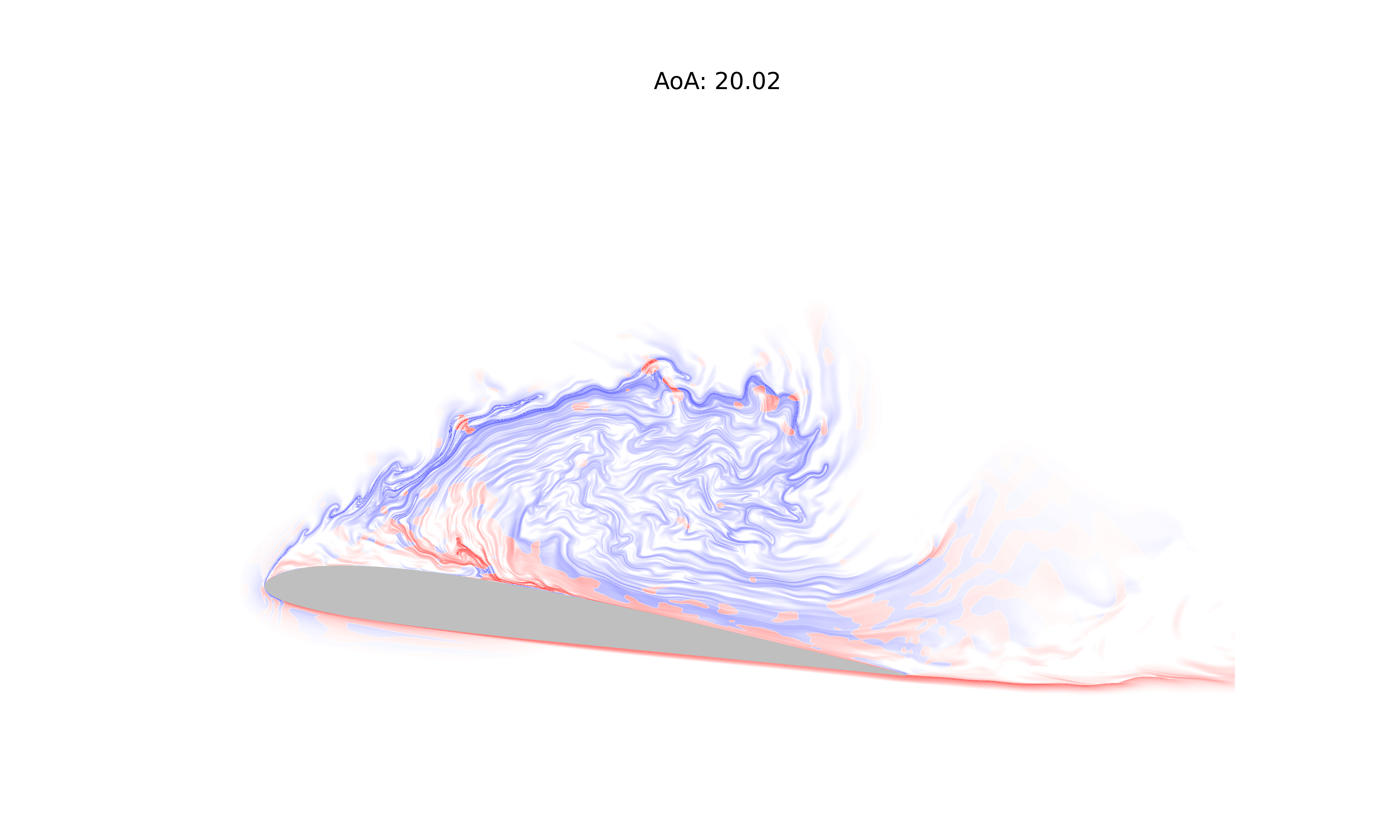}
        \put(0,15){(d)}
        \put(200,550){$AoA = 20^{\circ}$} 
    \end{overpic}
    \begin{overpic}[trim = 4.7cm 1cm 4cm 3cm,clip,width=0.32\textwidth]{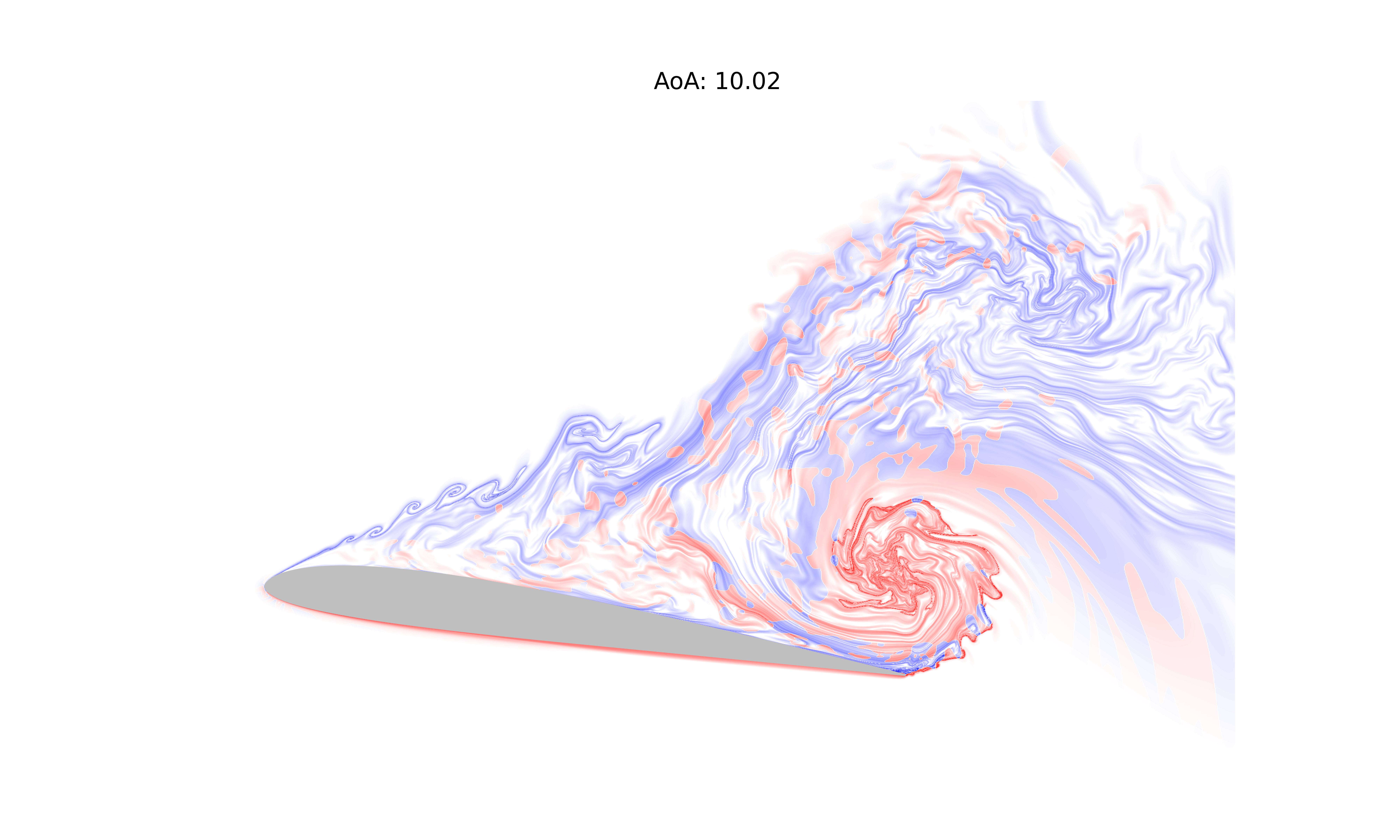}
        \put(0,15){(e)}
        \put(200,550){$AoA = 10^{\circ}$} 
    \end{overpic}
    \begin{overpic}[trim = 4.7cm 1cm 4cm 3cm,clip,width=0.32\textwidth]{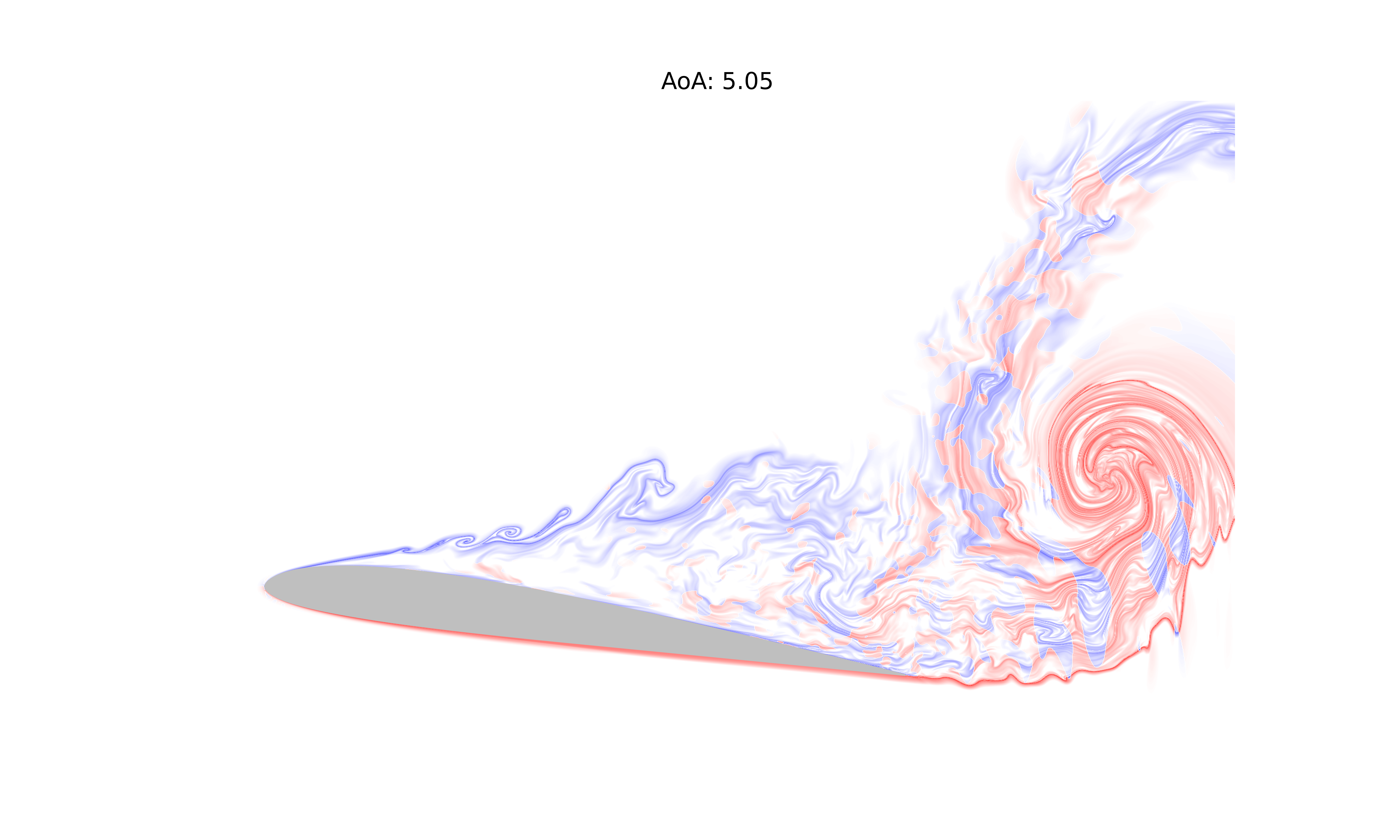}
        \put(0,15){(f)}
        \put(200,550){$AoA = 5^{\circ}$} 
    \end{overpic}
\caption{Evolution of the flow dynamics during the plunging cycle through visualization of $\text{FTLE}_{\Omega_{z}}$ fields.}
\label{fig : baseline}
\end{figure}

The contours of the negative FTLE fields are presented in Fig. \ref{fig : baseline} and also in Movie 1, submitted as supplementary material. This quantity provides a visualization of the flow structures in a similar fashion to a passive tracer. In order to permit a visualization of the DSV evolution, including the initial flow instabilities and the development of the trailing edge vortex (TEV), the spanwise vorticity $(\Omega_{z})$ contours are projected into the FTLEs as:

\begin{equation}
    \text{FTLE}_{\Omega_{z}} = \text{FTLE}(x,y,t) \cdot \text{sign}(\Omega_{z}(x,y,t)) \mbox{ ,}
\end{equation}
where

\begin{equation}
\text{sign}(\Omega_{z}) = 
\begin{cases} 
1, & \text{if } \Omega_{z} > 0 \\
0, & \text{if } \Omega_{z} = 0 \\
-1, & \text{if } \Omega_{z} < 0 \mbox{ .}
\end{cases}   
\end{equation}

The contours show the flow states at different phases of the motion cycle. During the downstroke, Kelvin-Helmholtz instabilities develop in the boundary layer while periodic vortex shedding can be observed at the trailing edge (Fig. \ref{fig : baseline}-a). The instabilities undergo a pairing process, being advected downstream while vorticity starts to accumulate at the leading edge (Fig. \ref{fig : baseline}-b). In the following instants, the DSV grows on the leading edge and begins to be transported, creating a zone of separated flow of the order of the airfoil chord (Figs. \ref{fig : baseline}-c and \ref{fig : baseline}-d). The passage of the DSV along the suction side of the airfoil induces a large increase in the drag coefficient, as will be seen later in Fig. \ref{fig : control_results}-a. Upon arriving at the trailing edge, the DSV causes a change in the circulation along the airfoil, giving rise to a TEV (\ref{fig : baseline}-e). The interaction between the two large-scale structures causes the ejection of the DSV from the airfoil surface. Then, when the airfoil resumes its upward movement, the TEV detaches from the surface and the boundary layer relaminarizes from the leading edge (Fig. \ref{fig : baseline}-f).

The ridges in the positive (negative) FTLE fields serve as analogs for stable (unstable) manifolds of the system. Therefore, the intersections of the ridges resemble saddle points for a dynamical system and, therefore, are regions of particular interest for flow control as they enable high amplification of disturbances. Thus, the identification of a potential saddle point in the flow enables one to pinpoint the areas which are more prone to respond to the blowing-suction actuation. With that, local stability analysis can be utilized to ascertain the optimal frequency and timing to actuate, achieving an improved aerodynamic performance by changing the flow topology.

\begin{figure}[H]
    \centering
    \begin{overpic}[trim = 0cm 0cm 2cm 0cm,clip,width=0.39\textwidth]{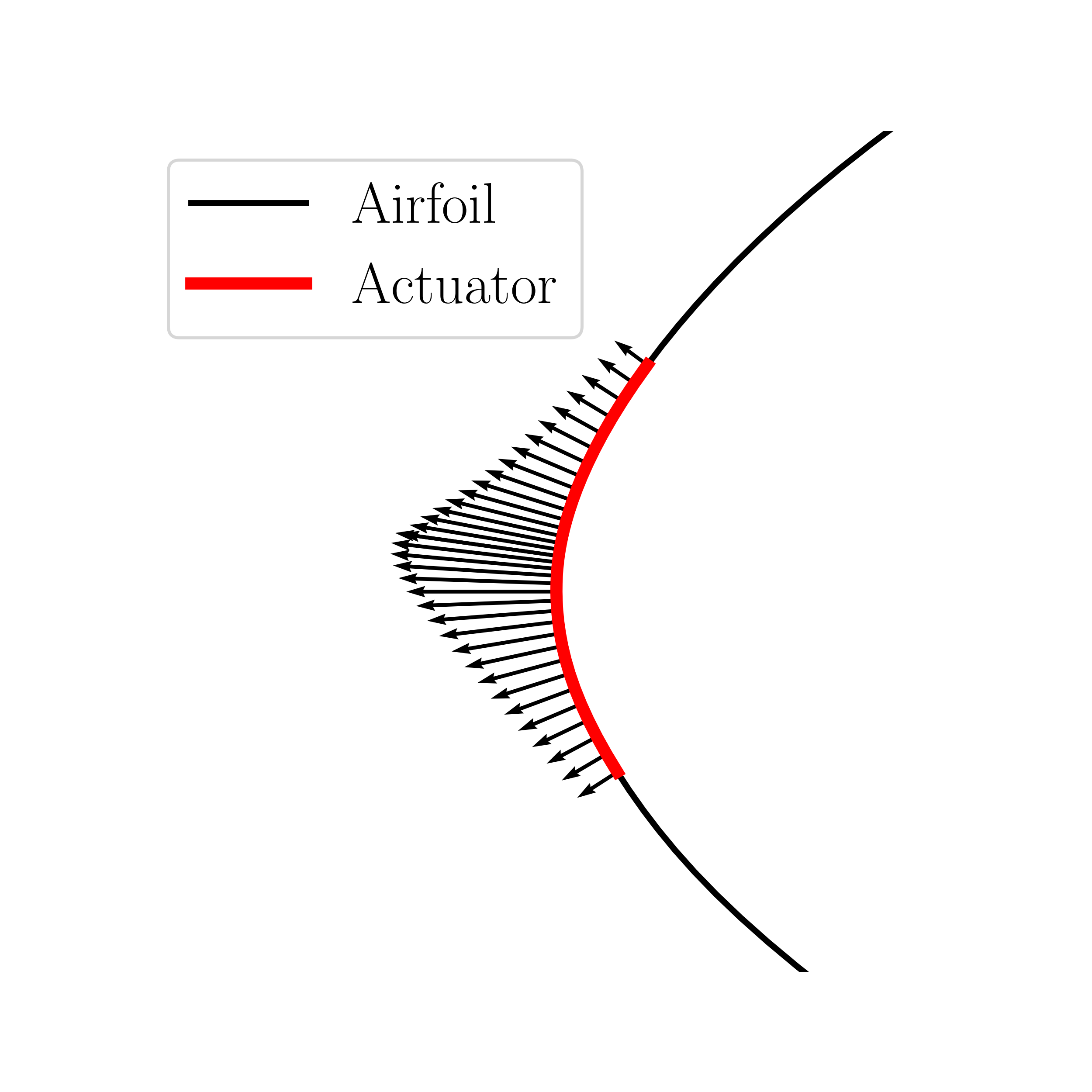}
        \put(0,100){a)}
    \end{overpic}
    \begin{overpic}[trim = 2cm 0cm 0cm 0cm,clip,width=0.39\textwidth]{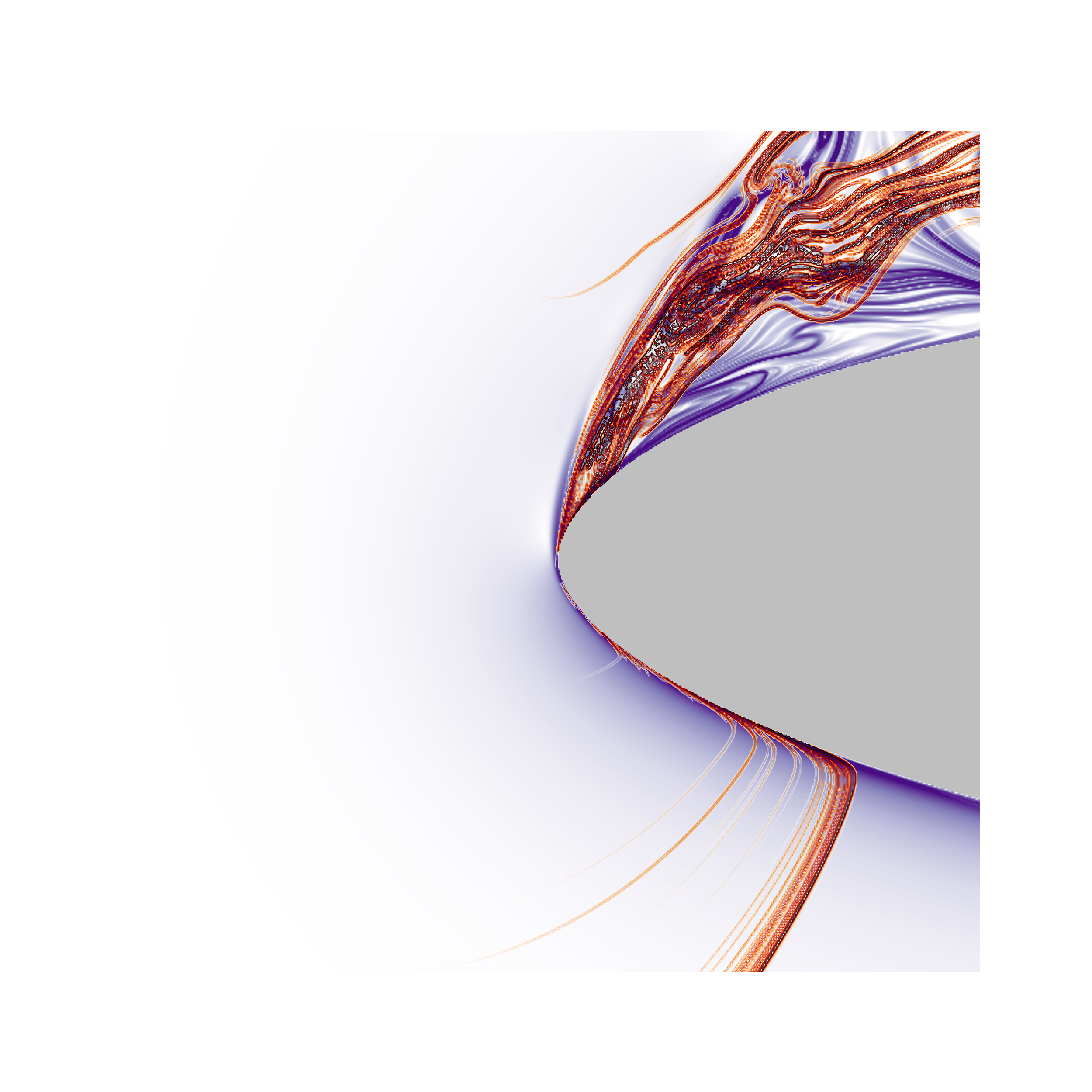}
        \put(0,100){b)}
    \end{overpic}
\caption{a) Placement of the suction and blowing actuator on the airfoil leading edge, and b) visualization of positive (orange) and negative (purple) FTLE contours at $AoA = 22^\circ$, emphasizing the saddle point dynamics near the leading edge}
\label{fig : saddle point}
\end{figure}

Figure \ref{fig : saddle point} presents an overlay of the positive (orange) and negative (purple) FTLE contours at $AoA = 22^\circ$, emphasizing the dynamics near the leading edge. Both contours are plotted in logarithmic scale. The positive FTLE contours highlight the stagnation point of the airfoil on the pressure side. This region depicts high sensitivity to small disturbances which can cause fluid particles to migrate to either side of the airfoil. However, the most pronounced sensitivity to disturbances is observed along the shear layer emanating on the suction side. Furthermore, we observe that the shear layer forms in a zone of intersection between the positive and negative FTLE contours, characterizing a potential saddle point of the flow. Similar observations have been made in the literature for a pitching airfoil \cite{Mulleners2012}. Therefore, we compute velocity profiles in this region to conduct stability analyses at each time instant to identify the most unstable frequencies and the optimal time window for flow actuation. It is also important to mention that, by trial and error, \citeauthor{Ramos_PRF} \cite{Ramos_PRF} observed that this particular region of the leading edge provided the best results in terms of flow response to the present blowing-suction actuation.

\subsection{Stability analysis}\label{subsec2}

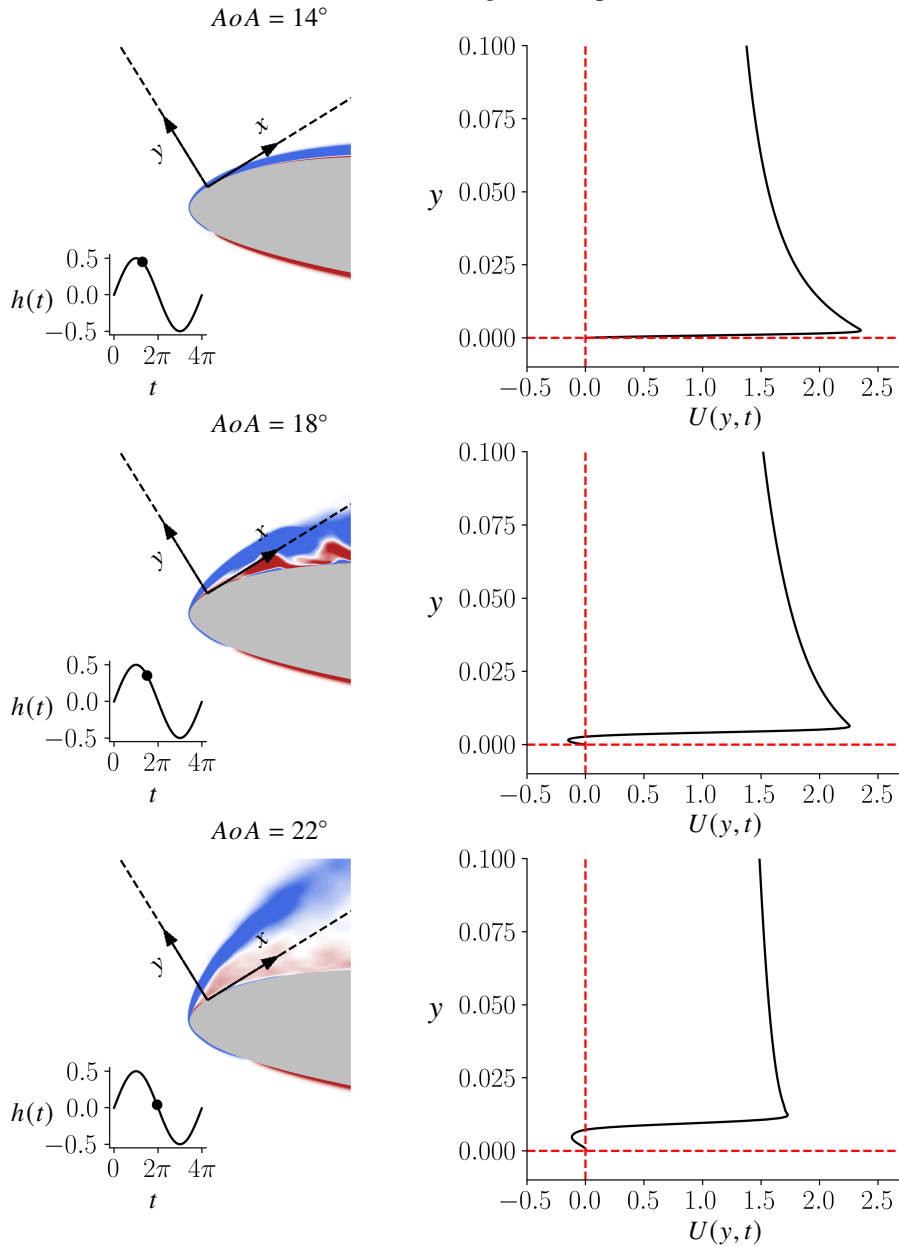
\begin{figure}[H]
    \centering
    \begin{overpic}[trim = 0cm 0cm 0cm 0.7cm,clip,width=0.7\textwidth]{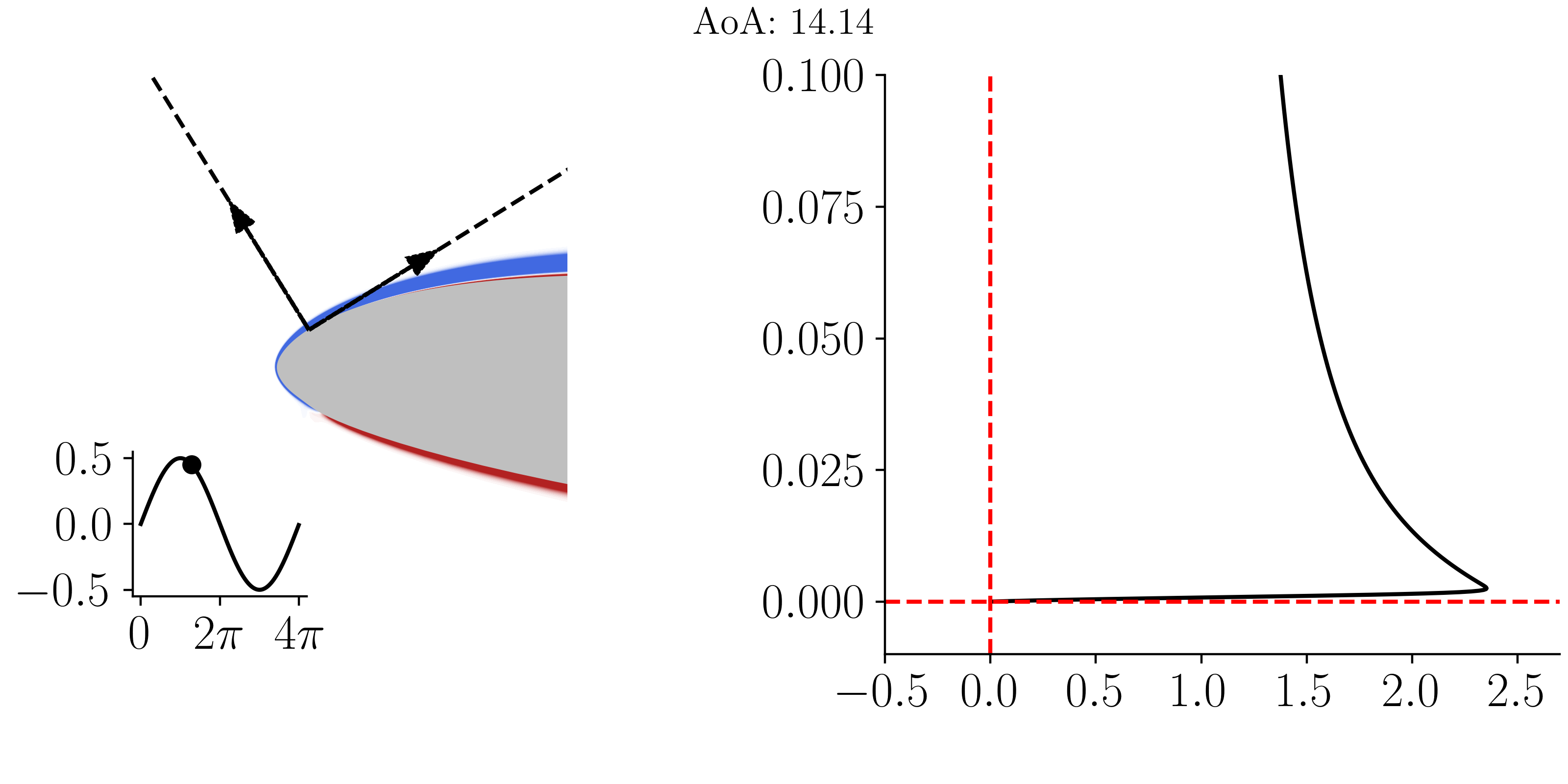}
        \put(-30,150){$h(t)$}
        \put(130,50){$t$}
        \put(200,475){$AoA = 14^\circ$}
        \put(450,270){\large \rotatebox{0}{$y$}}
        \put(750,15){$U(y,t)$}
        \put(130,320){\rotatebox{35}{$y$}}
        \put(250,350){\rotatebox{35}{$x$}}
    \end{overpic} 
     \begin{overpic}[trim = 0cm 0cm 0cm 0.7cm,clip,width=0.7\textwidth]{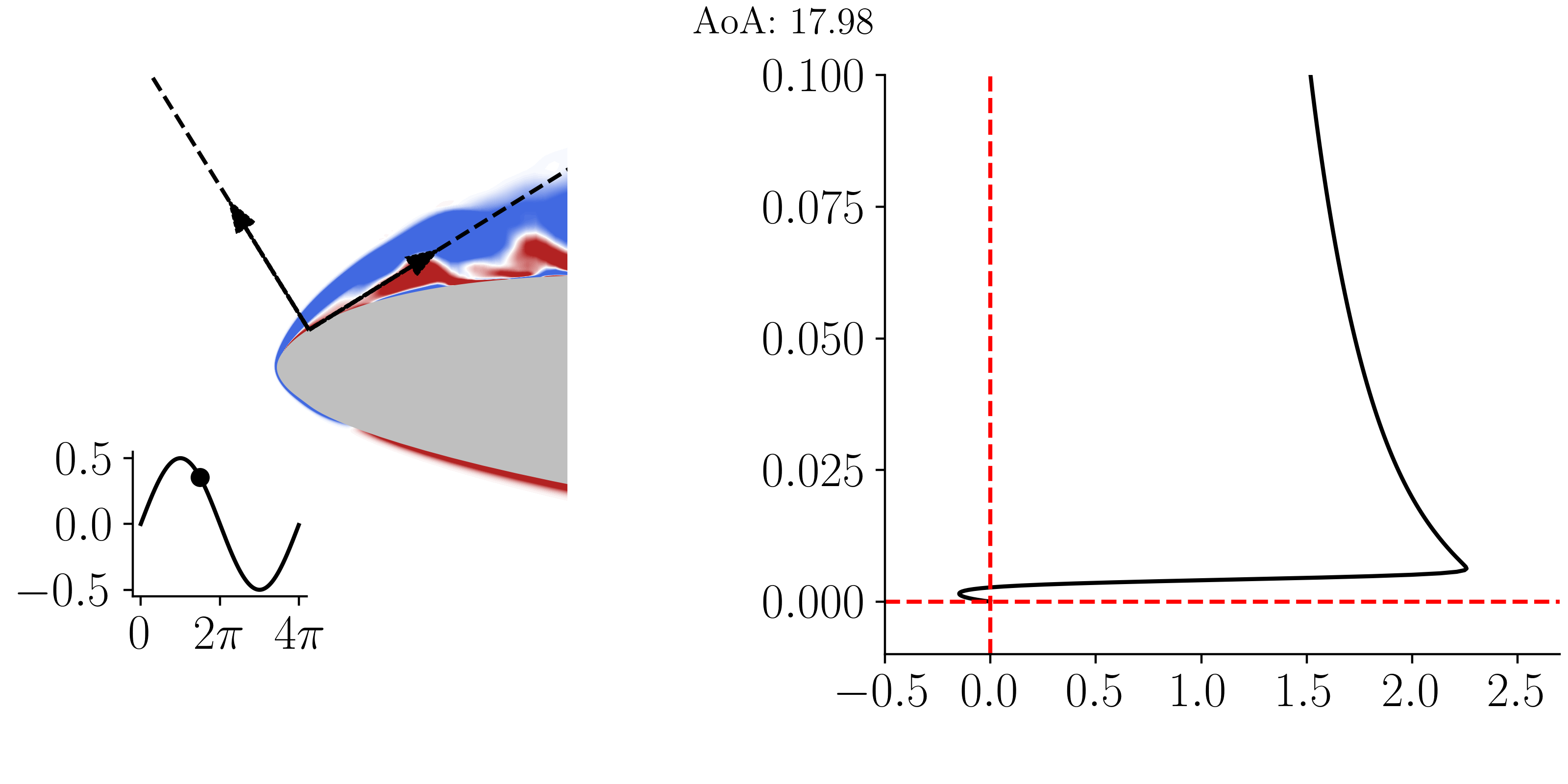}
        \put(-30,150){$h(t)$}
        \put(130,50){$t$}
        \put(200,475){$AoA = 18^\circ$}
        \put(450,270){\large \rotatebox{0}{$y$}}
        \put(750,15){$U(y,t)$}
        \put(130,320){\rotatebox{35}{$y$}}
        \put(250,350){\rotatebox{35}{$x$}}
    \end{overpic} 
    \begin{overpic}[trim = 0cm 0cm 0cm 0.7cm,clip,width=0.7\textwidth]{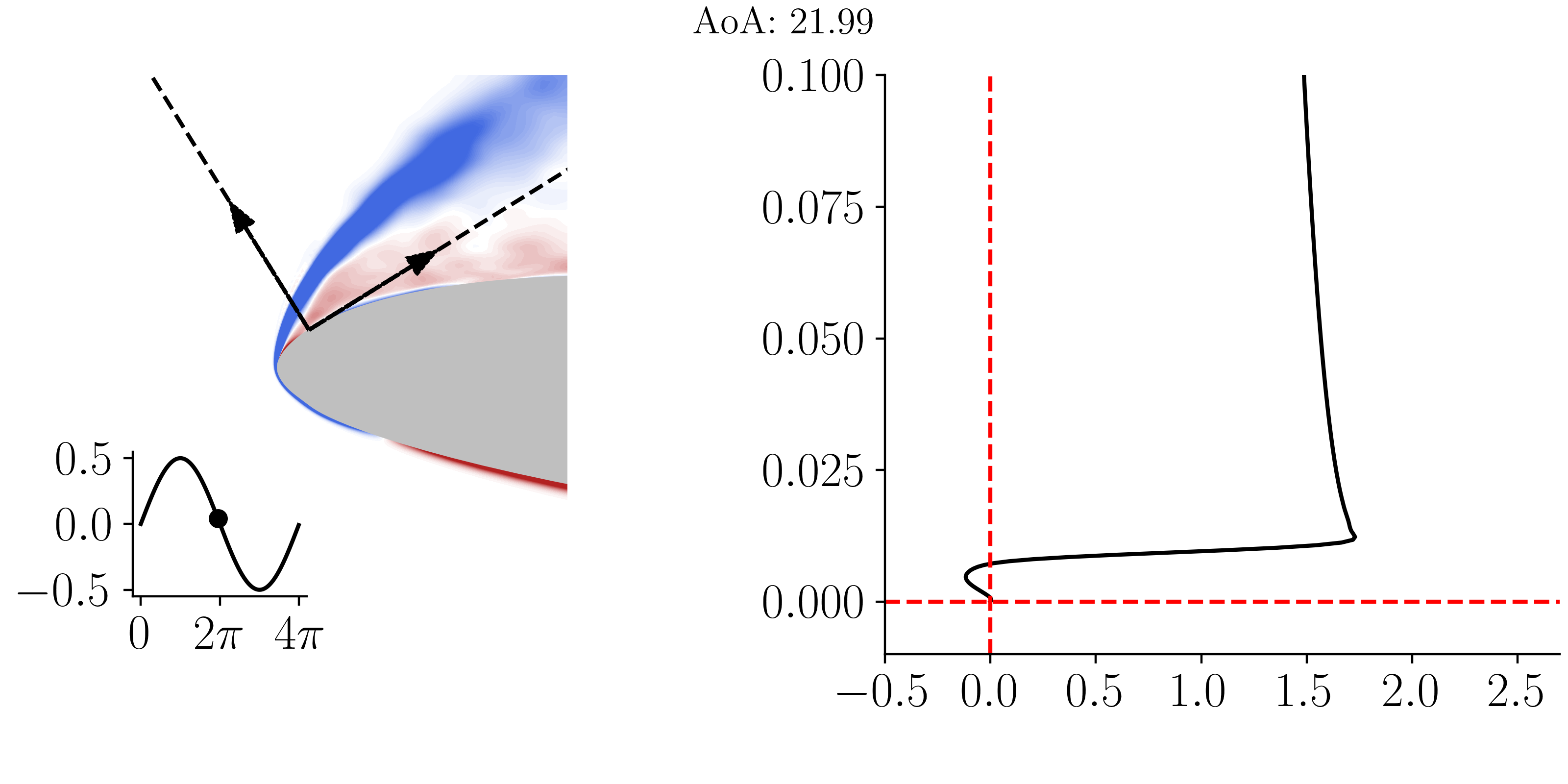}
        \put(-30,150){$h(t)$}
        \put(130,50){$t$}
        \put(200,475){$AoA = 22^\circ$}
        \put(450,270){\large \rotatebox{0}{$y$}}
        \put(750,15){$U(y,t)$}
        \put(130,320){\rotatebox{35}{$y$}}
        \put(250,350){\rotatebox{35}{$x$}}
    \end{overpic}     
\caption{Phase-averaged vorticity contours (left column) and tangential velocity profiles (right column). The black dashed lines show the location where the phase-averaged velocity profiles are computed, and the insets show the instant of the airfoil in the plunging cycle.}
\label{fig : profiles 1}
\end{figure}

In the previous section, the FTLE contours revealed that the flow dynamics in the airfoil leading edge region are relevant for the present study. As observed in Fig.~\ref{fig : baseline}, the shear layer developing at the leading edge is responsible for the DSV vorticity feeding. Therefore, a local stability analysis is conducted at that position about a series of phase-averaged velocity profiles as base flow. The phase-averaged base flow ($q_o = (U(y,t), 0, 0, 0)$) profiles are computed for 5 plunging cycles, being extracted along the wall-normal direction $(y)$ of the airfoil as illustrated in Fig.~\ref{fig : profiles 1}. The images show the evolution of $U(y,t)$ for different instants (effective $AoAs$) of the cycle. The left columns also depict the phase-averaged vorticity fields and the system of reference where the profiles are extracted. The second column shows the phase-averaged velocity profiles in the near wall region. In the vorticity plots, an auxiliary curve is provided to inform the instant of the airfoil in the plunging cycle. Although the figures show the velocity profiles only up to approximately $10\%$ of the chord in the wall-normal direction, the total length of the domain considered for the stability analysis corresponds to 50 chords in order to avoid confinement effects imposed on the eigenfunctions by farfield boundary conditions. As expected, the leading edge region is susceptible to considerable flow acceleration, which increases the local Mach number \cite{Miotto_Wolf_Gaitonde_Visbal_2022}. In order to verify the accuracy of the present incompressible linear operator, we compare its eigenspectrum with that computed by a compressible formulation. In the Appendix \ref{app:B}, a good agreement is found between both approaches, showing that compressibility effects do not change the present results and conclusions.

Figure \ref{fig : profiles 1} shows that during the initial instant of the downstroke, the boundary layer evolves under a favorable pressure gradient, evidenced by the flow acceleration near the wall compared to the freestream velocity. As the effective angle of attack increases, a reversed flow begins to develop, leading to the onset of an unsteady separation phenomenon as described by \citeauthor{Miotto_Wolf_Gaitonde_Visbal_2022} \cite{Miotto_Wolf_Gaitonde_Visbal_2022}. This process is shown by the vortical structures downstream of the location where the linear analysis is conducted at $AoA = 18^\circ$. As the airfoil continues its descent motion, we observe that the reversed flow region increases, leading to the detachment of the shear layer from the leading edge. In the phase-averaged vorticity contours, one can notice that the flow remains laminar throughout this period of the downward motion.

The stability analysis is conducted for the velocity profiles during the downstroke. For each instant, we sweep through streamwise wavenumbers ranging from $\alpha = 10 - 1000$ so that disturbances of length scales 0.62 to 0.0062 are considered. Both the wavenumbers and length scales are nondimensionalized by the airfoil chord. Only two-dimensional disturbances are considered ($\beta = 0$). This choice is justified by the results reported by \citeauthor{Ramos_PRF} \cite{Ramos_PRF}, who demonstrate that for the present flow, two-dimensional actuation is the most effective.

\begin{figure}[H]
    \centering
    \begin{overpic}[trim = 0cm 0cm 0cm 0cm,clip,width=0.72\textwidth]{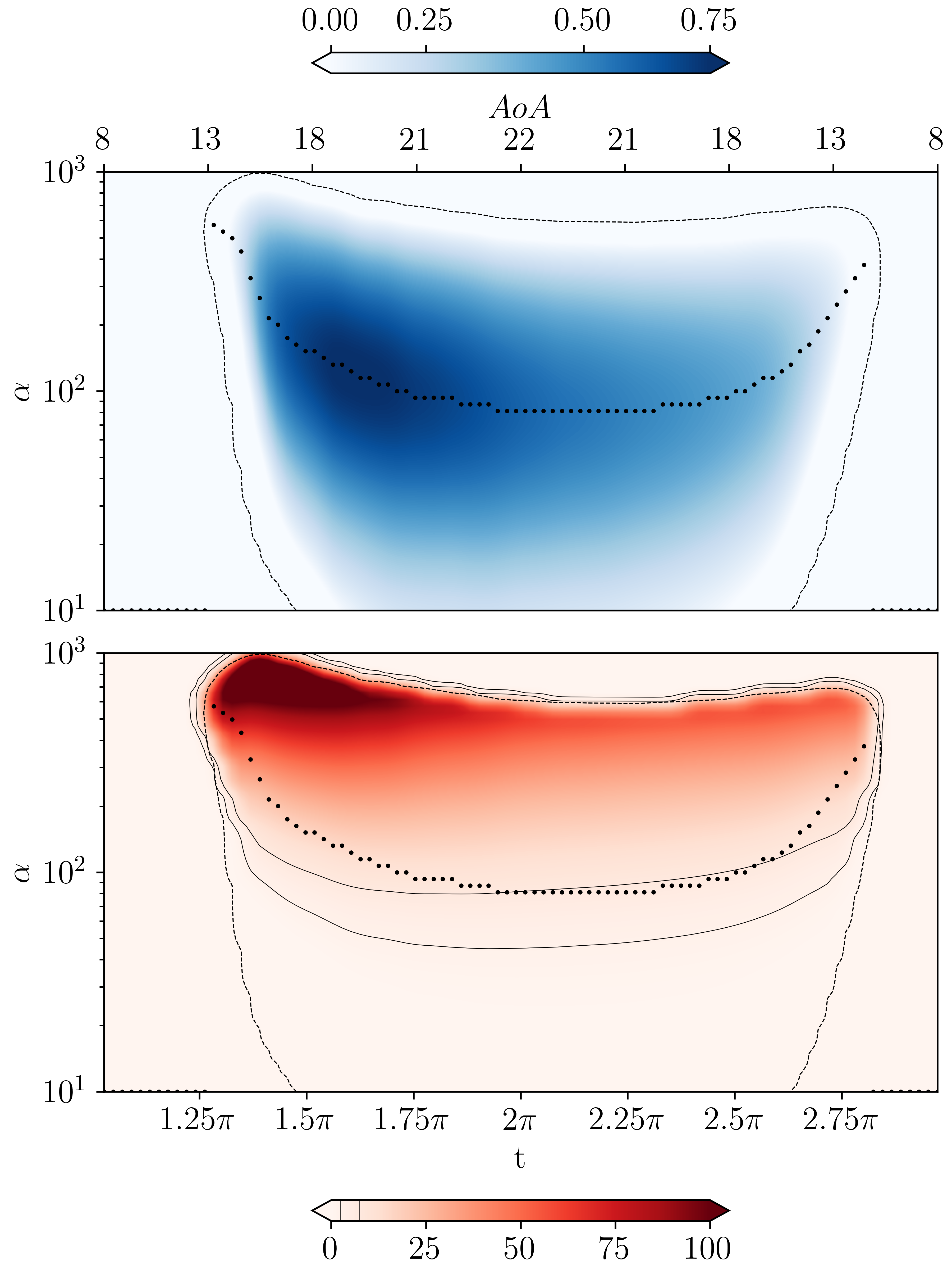}
        \put(90,800){\Huge Gr}
        \put(100,570){$Gr \leq 0$}
        \put(105,540){stable}
        \put(650,530){stable}
        \put(420,570){$Gr > 0$}
        \put(420,530){unstable}
        \put(420,230){$St= 2.5$}
        \put(420,310){$St= 7.5$}
        \put(90,400){\Huge St}
    \end{overpic}
\caption{Growth rate (top blue contours) and Strouhal number (bottom red contours) computed by the linear stability analysis for different streamwise wavenumbers $\alpha$ at different instants of the airfoil downward motion. Black dashed lines delimit the regions with a positive growth rate. In the bottom plot, isolines of $St = 2.5$ and $St = 7.5$ are also provided to highlight the regions of optimal frequencies reported by \citeauthor{Ramos_PRF} and the black dots show the most unstable eigenvalues as a function of time (lower $x$ axis) and $AoA$ (upper $x$ axis).}
\label{fig : stability contours}
\end{figure}
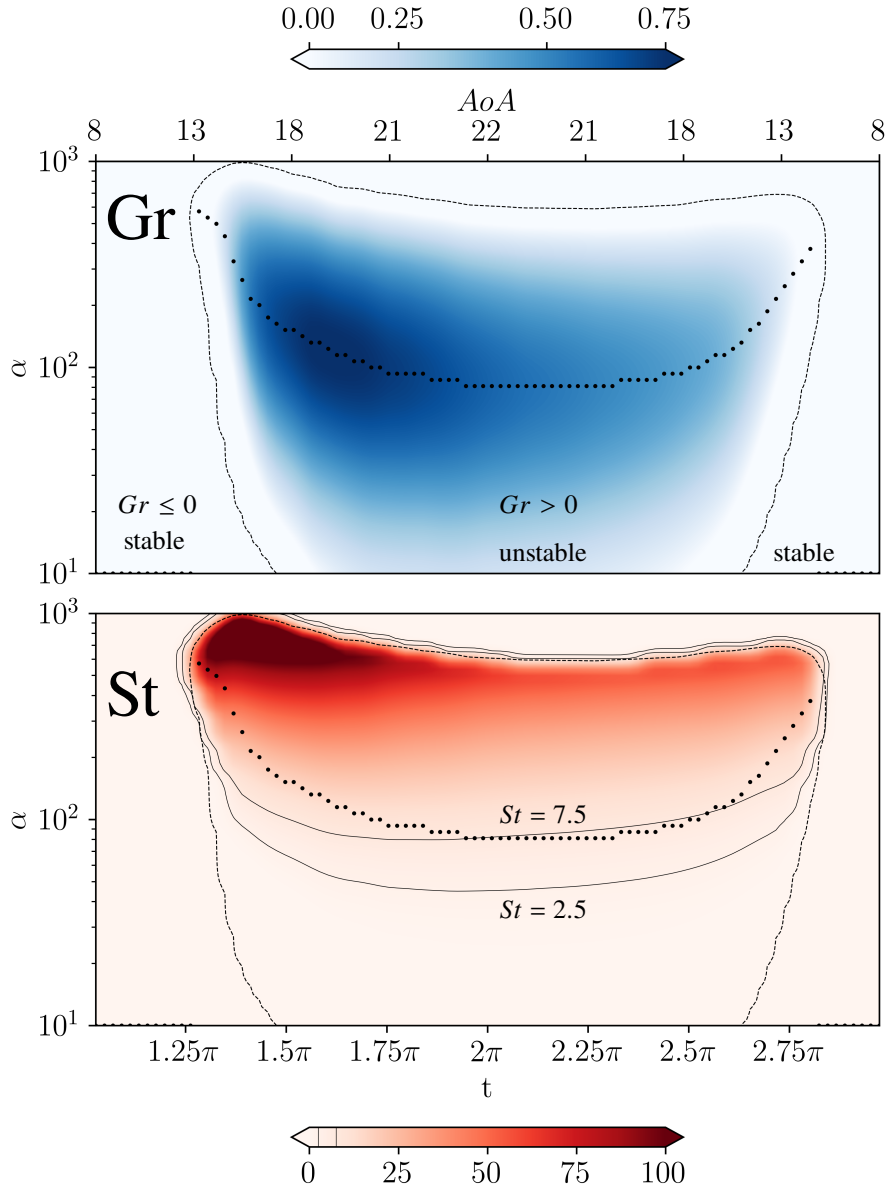

Results of the stability analysis are shown in Fig. \ref{fig : stability contours}, and the contours are generated by a method analogous to that used for computing a neutral stability curve. However, instead of varying the Reynolds number and the streamwise wavenumber, the contours are shown for different values of $\alpha$ as a function of time, i.e., for different phase-averaged velocity profiles computed during the airfoil motion. Thus, each point on the 2D contours is obtained by solving the eigenvalue problem in equation \ref{eq : eig} for a time instant (effective $AoA$) and a particular streamwise wavenumber. The most unstable eigenvalue from the spectrum is obtained and its growth rate and Strouhal number are plotted by the blue and red contours, respectively.

\begin{figure}[H]
    \centering
    \begin{overpic}[trim = 0cm 0cm 0cm 0cm,clip,width=0.99\textwidth]{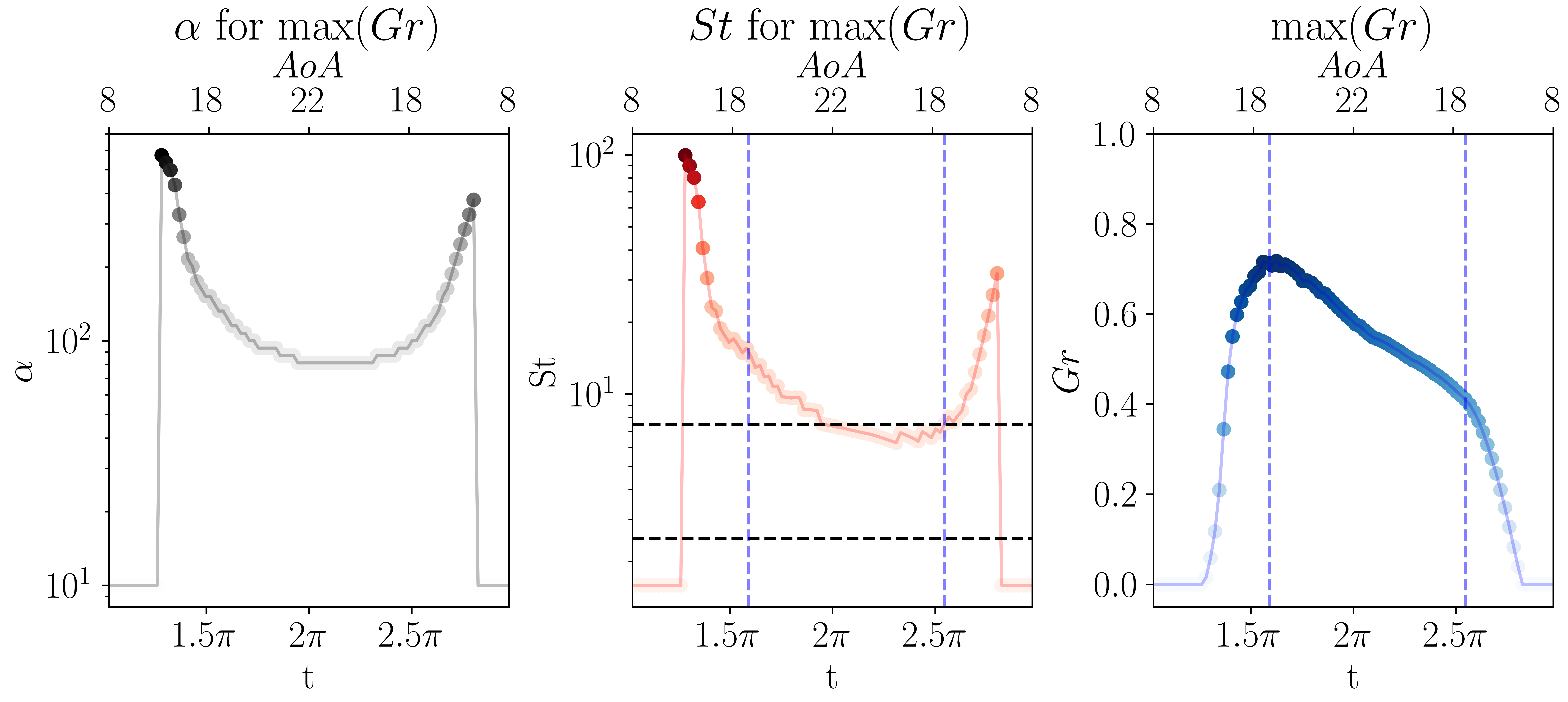}
        \put(20,0){$a)$}
        \put(360,0){$b)$}
        \put(700,0){$c)$}   
    \end{overpic} 
\caption{(a) Streamwise wavenumbers, (b) frequencies, and (c) growth rates of the most unstable eigenvalues at each time instant. Horizontal dashed lines depict the range of frequencies of optimal actuation observed by \cite{Ramos_PRF}, while the vertical dashed lines delimit the region of flow actuation suggested by the present stability analysis. Vivid colors are used for larger magnitudes in the plots.}
\label{fig : stability path}
\end{figure}

In the growth rate ($Gr$) contours we can see that the flow becomes unstable during the time $1.25 \pi \lesssim t \lesssim 2.75 \pi$. This is shown by the black dashed line that delimits the instant where a positive growth rate is computed from the spectrum. In the same contour, the black dots show the trajectory of the most unstable eigenvalues at each time instant. This trajectory provides the values of $\alpha$ that achieve maximum amplification at different instants of the downward motion, when the DSV forms. By plotting the eigenvalues over Strouhal number ($St$) contours, we obtain the corresponding frequencies for the unstable path. In the figure, isolines of $St = 2.5$ and $St = 7.5$ delimit the range of frequencies found by \citeauthor{Ramos_PRF} \cite{Ramos_PRF} to be  the best for actuation aiming drag reduction, obtained through a parametric study. Results for this most unstable trajectory are summarized in Fig. \ref{fig : stability path}, which shows the streamwise wavenumbers for the most unstable eigenvalues at each time instant, including also their frequencies and growth rates.

The temporal evolution of the parameters is shown in Fig. \ref{fig : stability path} for the most unstable eigenvalues.  We observe that, at early instants of the airfoil downward motion, the leading edge shear layer is unstable to small scale 2D disturbances (higher wavenumbers $\alpha$) and high frequencies. However, the growth rates of the high frequency eigenvalues reduce considerably with the downward motion, as shown in Fig. \ref{fig : stability contours}. Then, eigenmodes of larger wavelengths and lower frequencies become unstable. At $t \approx 2 \pi$, the streamwise wavenumber of the most unstable eigenvalue becomes constant for a longer time period, where $\alpha \approx 80$ and $\lambda_x \approx 8\%$ of the chord. Hence, disturbances imposed on this shear layer mode have more time to grow. This behavior of the eigenvalue spectra is expected due to the sinusoidal nature of the airfoil motion shown in Fig. \ref{fig : control_results}-b. As shown in this figure, for a reduced frequency $ \kappa = 0.5$, the variation in the effective angle of attack is minimal around $t = 2 \pi$ ($AoA = 22^\circ$). This result resonates with recent studies where stability analysis is applied to airfoils under small variations in the angle of attack \cite{Kern_Negi_Hanifi_Henningson_2024}. Hence, although in the present work the airfoil is subjected to an angle of attack variation in the range $-6^\circ \leq AoA \leq 22^\circ$, we can identify a region where $d (AoA) / d t \to 0$. 

\begin{figure}[H]
    \centering
    \begin{overpic}[trim = 0cm 0cm 0cm 0cm,clip,width=0.89\textwidth]{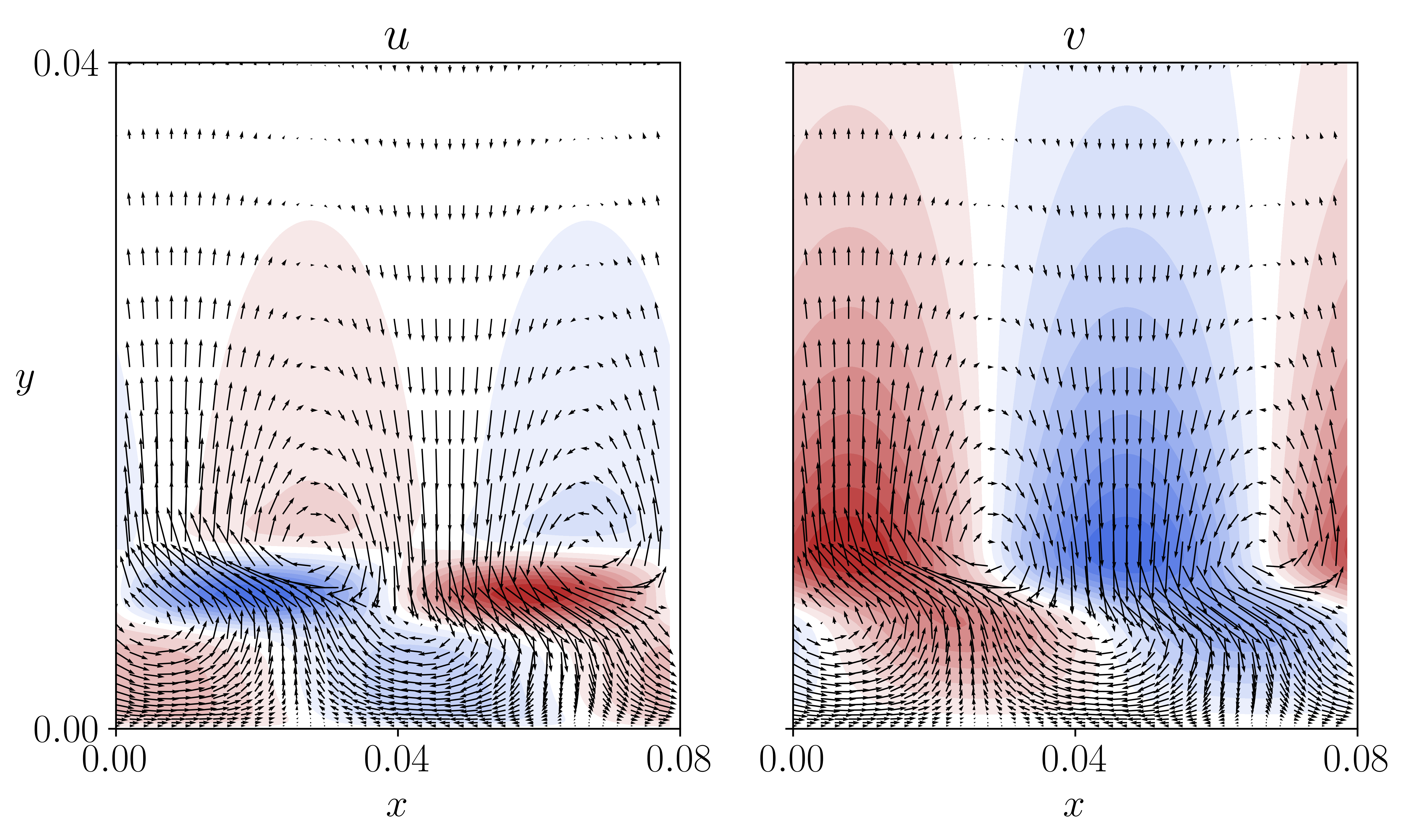}
    \end{overpic}
\caption{Velocity components $u$ and $v$ of the most unstable mode computed at $AoA = 22^\circ$ with $\alpha = 80$ depicting a Kelvin-Helmholtz instability. The spectrum for this particular case can be visualized in Fig. \ref{fig : validation}. in the Appendix \ref{app:B}}
\label{fig : unstable mode}
\end{figure}

For this particular region, the frequency of the most unstable eigenvalue becomes somewhat constant, with an approximate value of $St \approx 6.25$. It is worth mentioning that this frequency in within the range $2.5 \leq St \leq 7.5$ reported by the parametric investigation from \citeauthor{Ramos_PRF} \cite{Ramos_PRF} for the same flow configuration. This range is indicated by the isolines in Fig. \ref{fig : stability contours}, and by the horizontal black dashed lines in Fig. \ref{fig : stability path}. A two-dimensional visualization of the most unstable mode computed at $AoA = 22^\circ$ is provided in Fig. \ref{fig : unstable mode}. The spatial support of the mode appears along the shear layer, being a Kelvin-Helmholtz instability. The point of maximum tangential velocity coincides with the inflection point of the profile at $y \approx 0.008$.%

Results suggest that the constant frequency actuation of $St \approx 6.25$ is more efficient in disturbing the shear layer during the interval $2 \pi \leq t \leq 2.5 \pi$. During this time interval, the disturbances produced by the actuator are expected to develop into a Kelvin-Helmholtz instability with a size of approximately $8\%$ of the chord. This, in turn, should affect the development and evolution of the dynamic stall vortex. From this information we develop a control function which turns on the actuation only in the time interval informed by the linear stability analysis.

\subsection{Flow control}

\begin{figure}[H]
    \centering
    \begin{overpic}[trim = 0cm 0cm 0cm 0cm,clip,width=0.99\textwidth]{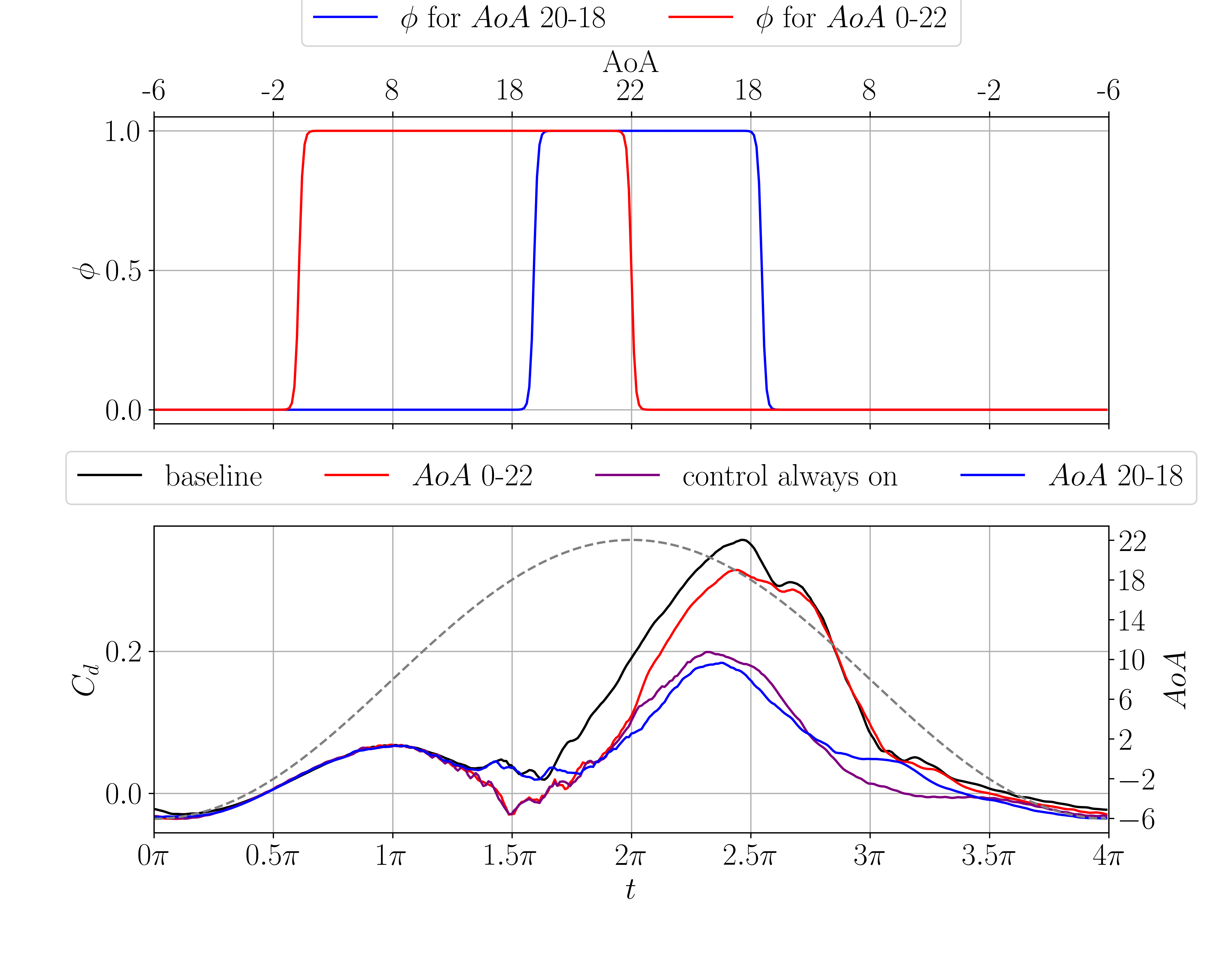}
        \put(0,520){(a)}
        \put(0,50){(b)}
    \end{overpic}
\caption{(a) Actuation function window $\phi$ as a function of time and effective $AoA$, and (b) phase-averaged drag coefficient $C_d$ (solid lines) as a function of time for different actuation setups and baseline (uncontrolled) flow. The evolution of the $AoA$ (dashed line) is also provided to highlight the fact that at $t = 2 \pi$ we have $d (AoA) / d t \to 0$}
\label{fig : control_results}
\end{figure}

In this section, we use the results from the stability analysis to inform the moment of the plunging cycle when the actuation should be turned on. Here, we employ the same actuation setup as in Ref. \cite{Ramos_PRF}, i.e., a periodic blowing and suction jet is applied at the airfoil leading edge. A 2D (spanwise) actuation is performed with a sinusoidal temporal function where the frequency is informed from the stability analysis. The jet velocity function is given by:
\begin{equation}
    \frac{U_{jet}}{U_{\infty}} (s,t) = \phi(t)\frac{U_{\text{jet max}}}{U_{\infty}}  \exp\left({\frac{-(s^* - 0.01)^2}{4.5}}\right) \sin{(St 2 \pi t)} \mbox{ ,} 
\end{equation}
where $s$ denotes the spatial location along the leading side of the airfoil and $s^* = 5(s - 0.005)$. A hyperbolic tangent function is used to create a coefficient $0 \leq \phi \leq 1$ that changes the amplitude of the jet according to the time in the cycle as:
\begin{equation}
 \phi =  \frac{e^{\nu(t -\gamma_1)}}{1+e^{\nu(t-\gamma_1 )}}\ -\frac{e^{\nu(t-\gamma_2)}}{1+e^{\nu(t-\gamma_2)}} \mbox{ .} \label{eq : control}
\end{equation}
Here, the factor $\nu$ delimits the slope of the transition region between the on and off states of the control, and the coefficients $\gamma_1$ and $\gamma_2$ dictate the instants to turn on and off the actuation, respectively. Based on the results of the stability analysis, we choose the value of $\gamma_1 = 1.65\pi$ to coincide with the peak moment of the growth rate informed by the stability analysis (see rightmost plot of Fig. \ref{fig : stability path}). The value  $\gamma_2 = 2.55\pi$ is chosen according to the moment when the frequency of the most unstable eigenvalue leaves the optimal actuation range of $2.5 \leq St \leq 7.5$. This interval of 2.82 convective time units corresponds to approximately 45\% of the airfoil descent time, or 22.5\% of the total cycle. In terms of the effective angle of attack, this time interval corresponds to turning the actuator on at $AoA = 20^\circ$ and turning it off at $AoA = 18^\circ$. This range is shown by the blue line in Fig. \ref{fig : control_results}(a). For all cases, we set the blowing and suction actuator frequency to $St = 6.25$ informed by the results of the stability analysis.

To demonstrate the assertiveness of the present approach, we compare the control results informed by the stability analysis with those from an actuation setup which is turned on mostly outside of the optimal temporal range, shown by the red line in Fig. \ref{fig : stability path}.
For this latter case, the function $\phi$ has values of $\gamma_1 = 0.62 \pi$ and $\gamma_2 = 2 \pi$ as shown in Fig. \ref{fig : control_results}(a). In the figure, the effective angle of attack is also provided in the top $x$ axis to show the moment in the plunging cycle when the control is active or inactive.

Figure \ref{fig : control_results}(b) presents the phase-averaged drag coefficient for the different cases considered. In these plots, the blue line identifies the results obtained with the control informed by the stability analysis. The red line indicates the case where actuation is turned on during a portion of the upstroke motion ($0^\circ\leq AoA \leq 8^\circ$ or $0.62 \pi \leq t \leq \pi$) and for half of the downstroke ($8^\circ\leq AoA\leq 22^\circ$ or $\pi \leq t \leq 2\pi$). The purple line depicts the drag coefficient obtained by keeping the actuator on during the entire cycle, while the black line shows the drag coefficient of the flow without actuation. 

Figure \ref{fig : control_results} shows that the drag coefficient remains at the same level for all cases until $t < 1.25\pi$. From this instant onwards, the red and purple drag curves separate from the other cases. This happens because the control is on for these particular simulations. At $t \approx 1.65\pi$, a second variation is observed for the blue line, and the drag coefficient for this setup (control turned on for $1.65 \pi< t < 2.55 \pi$) separates from that of the baseline case. The drag for all actuated cases follow the same trends for a short period ($ 1.65 \pi \leq t \leq 2\pi$), until the control is turned off for the less efficient setup (red line). From this moment, we observe that the peak drag of the ``optimal'' actuation case (blue) is similar to that where the control is always on (purple). The less efficient actuation setup (red line), despite presenting a reduction compared to the baseline flow, still exhibits a  higher drag peak than the more efficient case (blue line).

\begin{figure}[H]
    \centering
    \begin{overpic}[trim = 0cm 1.5cm 0cm 1.5cm,clip,width=0.99\textwidth]{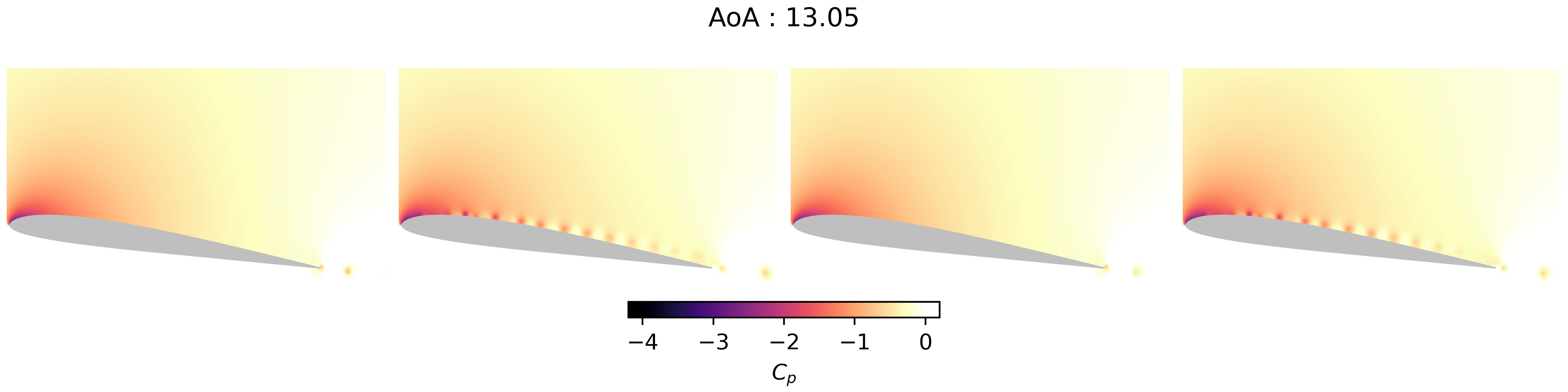}
        \put(75,150){baseline}
        \put(300,150){\textcolor{red}{$AoA=0^{\circ}-22^{\circ}$}}
        \put(550,150){\textcolor{blue}{$AoA=20^{\circ}-18^{\circ}$}}
        \put(800,150){\textcolor{purple}{always on}}
        \put(-15,20){\rotatebox{90}{\footnotesize $AoA \approx 13^{\circ}$}}
        \put(1000,20){\rotatebox{90}{\small $t \approx 1.22 \pi$}}
        \put(15,10){\small off}
        \put(260,10){\small on}
        \put(510,10){\small off}
        \put(760,10){\small on}
    \end{overpic} 
    \begin{overpic}[trim = 0cm 1.5cm 0cm 1.5cm,clip,width=0.99\textwidth]{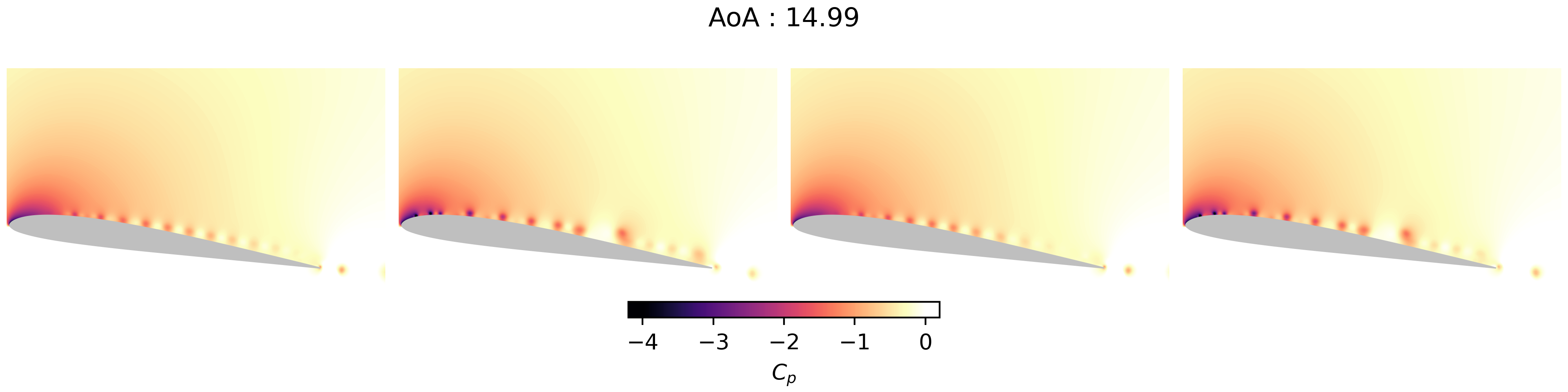}
        \put(-15,20){\rotatebox{90}{\footnotesize$AoA \approx 15^{\circ}$}}
        \put(1000,20){\rotatebox{90}{\small $t \approx 1.32 \pi$}}
        \put(15,10){\small off}
        \put(260,10){\small on}
        \put(510,10){\small off}
        \put(760,10){\small on}
    \end{overpic}
    \begin{overpic}[trim = 0cm 1.5cm 0cm 1.5cm,clip,width=0.99\textwidth]{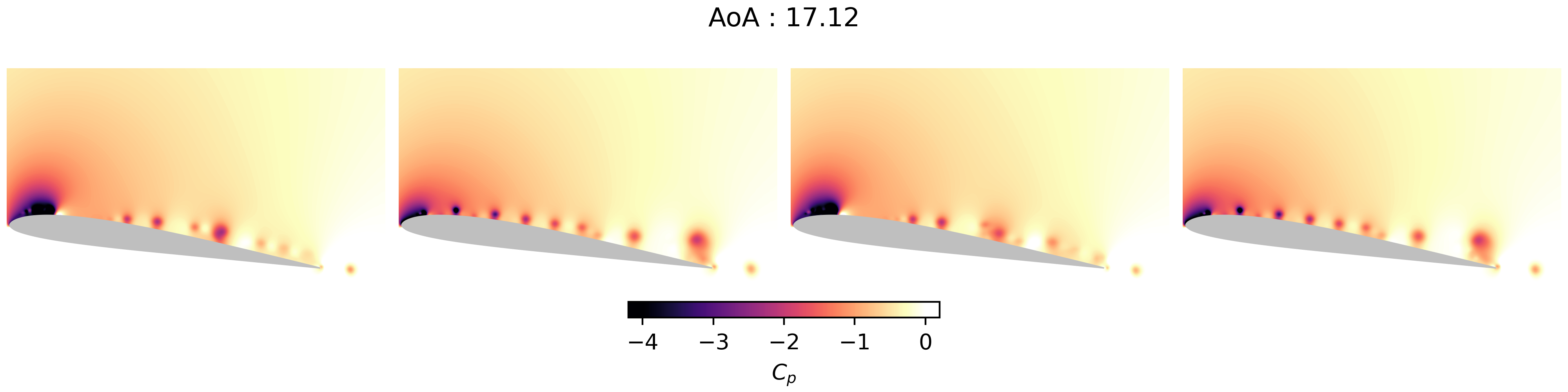}
        \put(-15,20){\rotatebox{90}{\footnotesize$AoA \approx 17^{\circ}$}}
        \put(1000,20){\rotatebox{90}{\small $t \approx 1.44 \pi$}}
        \put(15,10){\small off}
        \put(260,10){\small on}
        \put(510,10){\small off}
        \put(760,10){\small on}
    \end{overpic}
    \begin{overpic}[trim = 0cm 1.5cm 0cm 1.5cm,clip,width=0.99\textwidth]{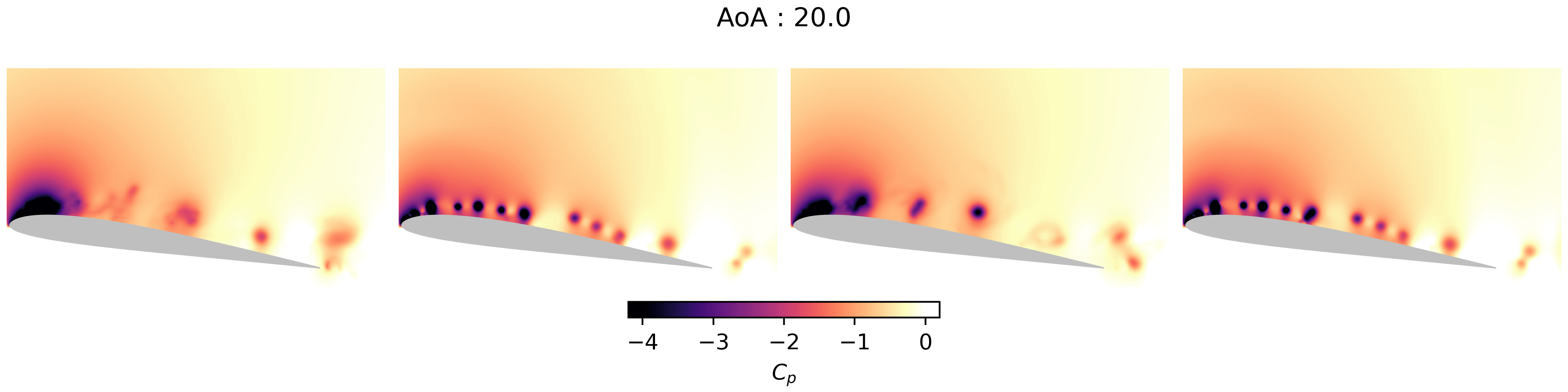}
        \put(-15,20){\rotatebox{90}{\footnotesize$AoA\approx 20^{\circ}$}}
        \put(1000,20){\rotatebox{90}{\small $t \approx 1.65 \pi$}}
        \put(15,10){\small off}
        \put(260,10){\small on}
        \put(510,10){\small on}
        \put(760,10){\small on}
    \end{overpic}
    \begin{overpic}[trim = 0cm 1.5cm 0cm 1.5cm,clip,width=0.99\textwidth]{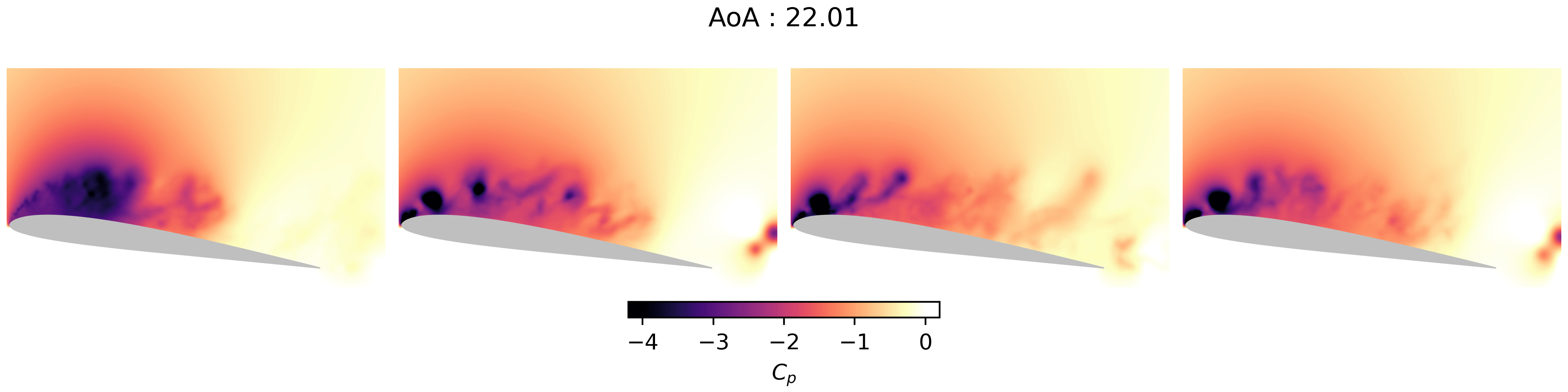}
        \put(-15,20){\rotatebox{90}{\footnotesize $AoA \approx 22^{\circ}$}}
        \put(1000,20){\rotatebox{90}{\small $t \approx 2 \pi$}}
        \put(15,10){\small off}
        \put(260,10){\small off}
        \put(510,10){\small on}
        \put(760,10){\small on}
    \end{overpic}
    \begin{overpic}[trim = 0cm 1.5cm 0cm 1.5cm,clip,width=0.99\textwidth]{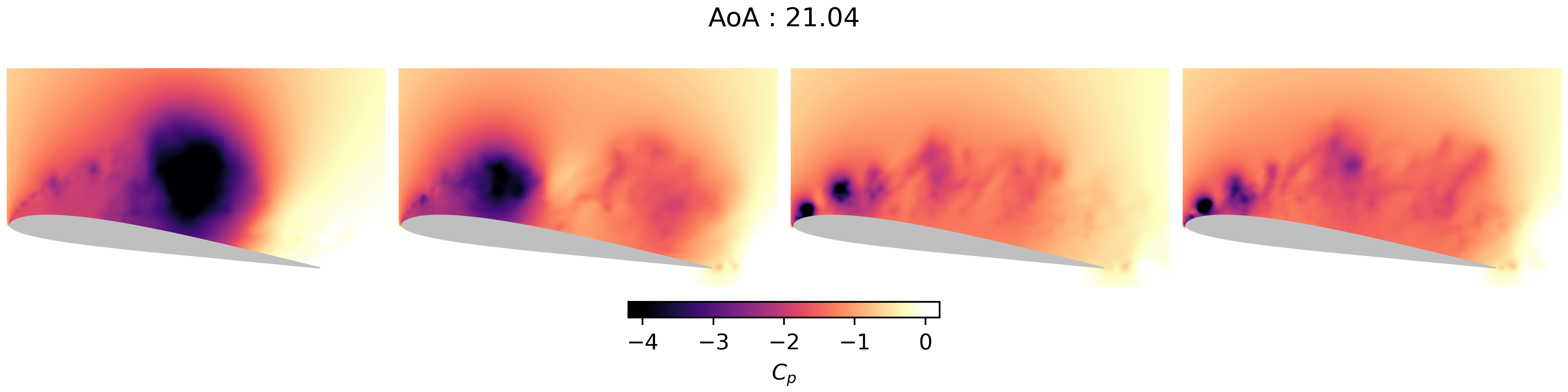}
        \put(-15,20){\rotatebox{90}{\footnotesize $AoA \approx 21^{\circ}$}}
        \put(1000,20){\rotatebox{90}{\small $t \approx 2.25 \pi$}}
        \put(15,10){\small off}
        \put(260,10){\small off}
        \put(510,10){\small on}
        \put(760,10){\small on}
    \end{overpic}
    \begin{overpic}[trim = 0cm 1.5cm 0cm 1.5cm,clip,width=0.99\textwidth]{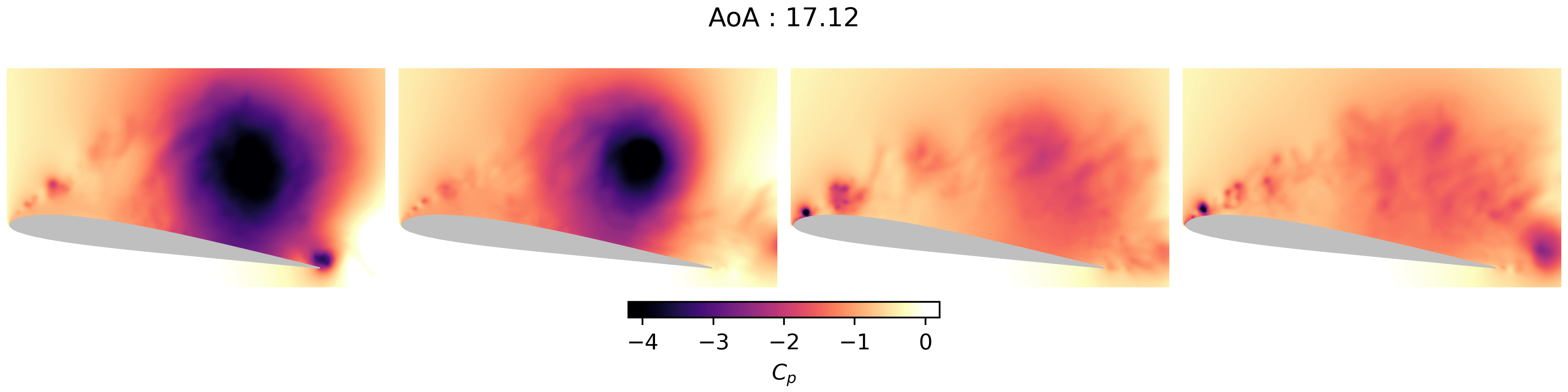}
        \put(-15,20){\rotatebox{90}{\footnotesize $AoA \approx 17^{\circ}$}}
        \put(1000,20){\rotatebox{90}{\small $t \approx 2.56 \pi$}}
        \put(15,10){\small off}
        \put(260,10){\small off}
        \put(510,10){\small off}
        \put(760,10){\small on}
    \end{overpic}
    \begin{overpic}[trim = 0cm 0cm 0cm 1.5cm,clip,width=0.99\textwidth]{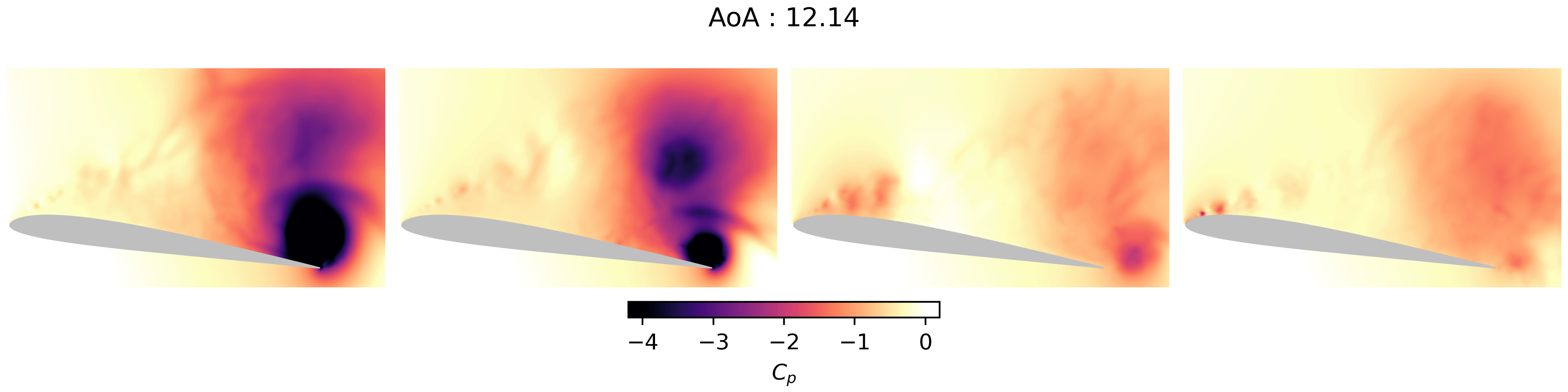}
        \put(-15,65){\rotatebox{90}{\footnotesize $AoA \approx 12^{\circ}$}}
        \put(1000,65){\rotatebox{90}{\small $t \approx 2.82 \pi$}}
        \put(15,65){\small off}
        \put(260,65){\small off}
        \put(510,65){\small off}
        \put(760,65){\small on}
    \end{overpic}
\caption{Evolution of instantaneous pressure coefficient $C_p$ contours (top to bottom) for the 4 cases considered at selected angles of attack during the downstroke motion. From left to right: baseline (uncontrolled) case, less efficient case with control turned on during $0^\circ \leq AoA \leq 22^\circ$, stability-analysis-informed case, and case where the control is always on. The full evolution of the flow field can be visualized in movie 2 submitted as supplementary material.}
\label{fig : contro_states} 
\end{figure}
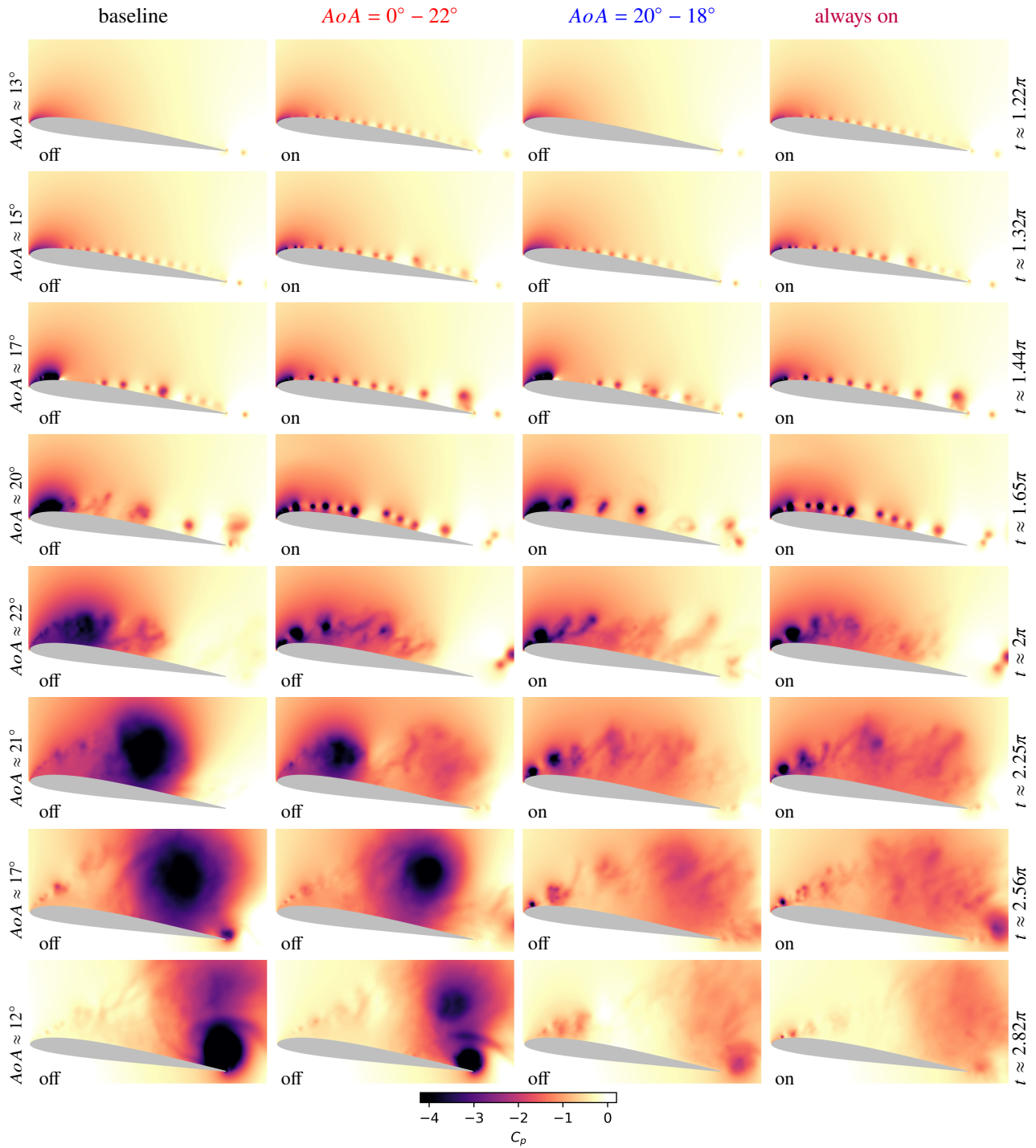

The pressure coefficient $C_p$ contours shown in Fig. \ref{fig : contro_states} provide a better understanding of the flow response to the present actuation setups. From left to right, the panels present the instantaneous $C_p$ flowfields for the baseline case, the less efficient actuation case, the stability-analysis-informed actuation case, and the configuration where control is turned on during the entire plunging cycle. The temporal evolution of $C_p$ is given by the sequence of panels from top to bottom, with the effective attack angle indicated on the left hand side of the plots at each stage. A text is also provided in each individual plot to inform if the control is on or off for that particular moment.

The first row of the figure shows an instant in the beginning of the downward motion. The actuation results in the early onset of Kelvin-Helmholtz instabilities that develop on the suction side of the airfoil, as depicted by the second and fourth columns. For the other cases, the boundary layer remains undisturbed. The following row, for $AoA \approx 15^\circ$, shows a delay in the appearance of the instabilities, besides a difference in their wavelengths. Between $17^\circ \leq AoA \leq 20^\circ$, columns 1 and 3 show the formation of a coherent DSV connected to the leading edge, while the cases with actuation display the ejection of smaller scale vortices which prevent the accumulation of vorticity at the leading edge. As the airfoil motion continues for $AoA = 22^\circ$and $21^\circ$, the DSV forms and ejects in the baseline case, being transported by the flow. The second column shows that turning the actuation off at $AoA = 22^\circ$ also leads to formation of a DSV, although it is smaller compared to the baseline case. The other cases where the actuation are kept turned on still present smaller vortices which are ejected from the leading edge, without the formation of a continuous shear layer that feeds the growth of a coherent DSV. The sizes of these smaller scale vortices are around 8\% of the chord, being in agreement with the stability analysis prediction.

In the following instants, at $AoA = 17^\circ$, we see that the shear layer feeds the DSV for the cases depicted in columns 1 and 2. However, for the previously actuated flow in the second column, the DSV is slightly weaker. The actuated case of column 3 has an even weaker and more diffused DSV due to the flow control performed in the right timing, provided by the linear stability analysis. The result from column 3 is similar to that of column 4, for which the actuation is always turned on. The transport of the DSV is responsible for the major drag peaks shown in Fig. \ref{fig : control_results}(b), while the TEV is responsible for the secondary smaller peak following the DSV. This latter structure is absent from the actuated cases shown in columns 3 and 4, in the last row.

In summary, the success of the control strategy relies on suppressing the growth stage of the DSV by disturbing the shear layer that feeds it. The introduction of disturbances at the leading edge creates instabilities in the shear layer which cut off the continuous flow of vorticity to the DSV. This causes the ejection of smaller scale vortices due to actuation at the right timing computed from the linear stability analysis, which suppresses the generation of a coherent vortex structure.

\section{Conclusions}

Drag reduction in dynamic stall is achieved by a flow control strategy informed by linear stability analysis. The methodology is applied for an SD7003 airfoil undergoing a periodic plunging motion which depicts a deep dynamic stall condition. The unsteady flow is analyzed by visualization of positive and negative FTLE fields, which provide analogs for stable and unstable manifolds of the fluid system. The intersection of the FTLE ridges near the leading edge reveals a saddle point with potential for flow control application. A local stability analysis is conducted at this saddle point employing phase-averaged velocity profiles throughout the airfoil descent motion. It reveals the moment at which the flow becomes unstable due to a Kelvin-Helmholtz instability arising along the shear layer forming at the airfoil leading edge. Results show that the frequency and wavenumber of the most unstable eigenvalues become nearly constant when the variation of the effective angle of attack is minimal. Moreover, this frequency is within the range of optimal actuation as reported by the parametric study from \citeauthor{Ramos_PRF} \cite{Ramos_PRF}.

Wall-resolved LES are performed to assess the potential of the flow control framework which is performed using the same blowing and suction actuation setup as in the above reference. However, in the present work, the actuation is turned on only over a finite time window informed by the stability analysis.
For this, a control function is applied taking the time interval and frequency as inputs when the unstable eigenvalues become nearly constant.
The control results using the reduced actuation duty cycle support the utility of the stability analysis. With a significant reduction of the duty cycle by 77.5\%, the drag is reduced by the same level as that observed in the case where the actuation is turned on over the entire plunging cycle. The comparison with a less efficient actuation case, where the actuator is kept on outside the optimal region informed by the stability analysis, reinforces the conclusions obtained by the present framework. Results from the LES show that the disturbances introduced by the actuation modify the leading-edge dynamics by interrupting the continuous accumulation of vorticity that feeds the DSV. Instead, discrete vortices with length scales predicted by the stability analysis are ejected and transported. These smaller scale structures are shown to induce a lower unsteady drag compared to the large-scale DSV. 

\section*{Appendix}
\subsection{Computation of finite-time Lyapunov exponents}\label{app:A}

The computation of the FTLE fields is based on the standard approach described by \citet{Brunton_LCS}. Initially, a grid of particles $X_{0} \subset \mathbb{R}^{2} $ is defined across the domain of interest. These particles are integrated with the velocity field from the initial time $0$ to the final time $T$, yielding a time-T particle flow map denoted as $\Phi_{0}^{T}$ defined as follows:

\begin{equation}
    \Phi_{0}^{T}: \mathbb{R}^{2} \to \mathbb{R}^{2}; \mathbf{X}(0) \mapsto \mathbf{X}(T)  \mbox{ ,}
\end{equation}
    
\begin{equation}
    \mathbf{X}(T) = \mathbf{X}(0) + \int_{0}^{T} \mathbf{u}(\mathbf{x}(\tau), \tau)  d\tau \mbox{ .}
\end{equation}
Here, $\mathbf{u}(\mathbf{x}(\tau), \tau)$ denotes the time-dependent velocity field over the particle trajectory $\mathbf{x}(\tau)$ at a time $\tau$. 

The flow map Jacobian $\mathbf{D} \Phi_{0}^{T}$ is then computed by a 2nd-order central finite difference scheme using the neighbouring particles in a Cartesian mesh such as:
\begin{equation}
\mathbf{D} \Phi_{0}^{T} = 
\begin{bmatrix}
\frac{\Delta x(T)}{ \Delta x (0)} & \frac{\Delta x(T)}{ \Delta y (0)} \\
\frac{\Delta y(T)}{ \Delta x (0)} & \frac{\Delta y(T)}{ \Delta y (0)} 
\end{bmatrix}
=
\begin{bmatrix}
\frac{x_{i+1,j}(T)- x_{i-1,j} (T) }{ x_{i+1,j}(0)- x_{i-1,j}(0) } &  \frac{x_{i,j+1}(T)- x_{i,j-1} (T) }{ y_{i,j+1}(0)- y_{i,j-1}(0)}\\
\frac{y_{i+1,j}(T)- y_{i-1,j} (T) }{ x_{i+1,j}(0)- x_{i-1,j}(0) } &  \frac{y_{i,j+1}(T)- y_{i,j-1} (T) }{ y_{i,j+1}(0)- y_{i,j-1}(0) } 
\end{bmatrix} \mbox{ .}
\end{equation}
In this Appendix, $x$ and $y$ denote the coordinates of the particles and the subscripts $i$ and $j$ relate to their indices in the computational domain. Finally, the Cauchy-Green deformation tensor is computed as:
\begin{equation}
    \mathbf{\Delta} = (\mathbf{D} \Phi_{0}^{T})^{\ast} \mathbf{D} \Phi_{0}^{T} \mbox{ ,}
\end{equation}
where $\ast$ denotes the transpose, and the largest eigenvalue $(\lambda_{max})$ from this tensor is computed to form the FTLE field
\begin{equation}
    \sigma (\mathbf{D} \Phi_{0}^{T} ; \mathbf{x_{0}}) = \frac{1}{\lvert T \rvert} \log{ \sqrt{\lambda_{max}(\Delta (\mathbf{x_{0}}))}} \mbox{ .}
\end{equation}

\begin{figure}[!h]
\begin{center}
\includegraphics[trim={1cm 2.0cm 1cm 2.5cm},clip, width=.95\textwidth]{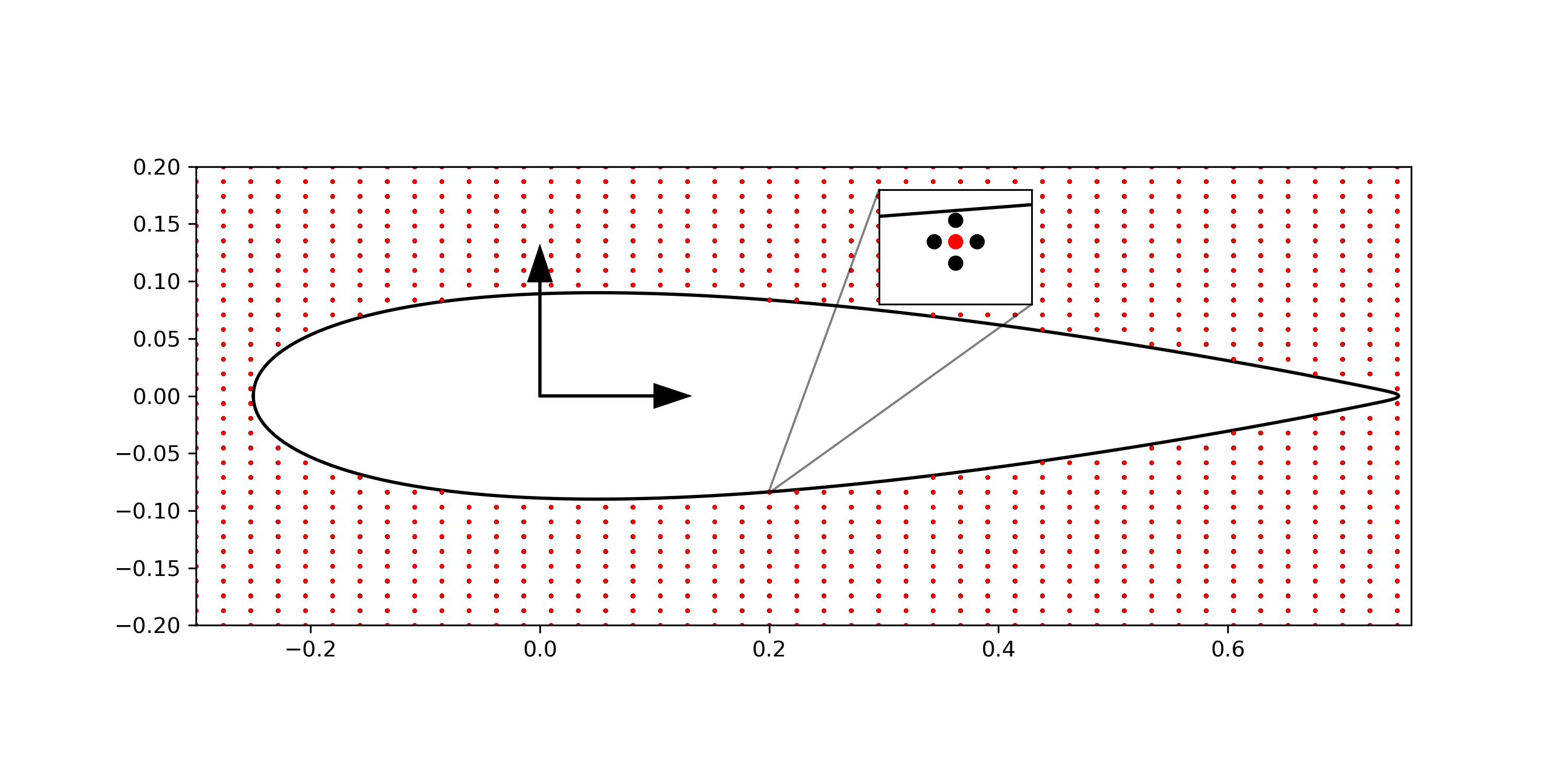}
\end{center}
\caption{Example of auxiliary grid used to compute the flow map Jacobian. The properties at the reference point marked in red are obtained by integrating and computing a 2nd-order central finite-difference scheme in the neighboring black particles.}
\label{fig : numerical method}
\end{figure}

A reference mesh,  indicated by the red dots in the figure, is first placed on top of the LES grid, being used to plot the FTLE fields. 
Then, the FTLEs are computed using an auxiliary grid (represented by the black dots in the inset) in which the LES properties are interpolated. Figure \ref{fig : numerical method} presents an example of the spatial discretization approach used in the present study. The auxiliary grid is constructed with the maximum distance allowed from the reference grid points to the airfoil surface, ensuring that all points remain outside of the solid body, as shown in the figure inset. The auxiliary points are then integrated along the flow using a second-order Adams-Bashforth method. At each integration step, the velocity field in the position of the particles is determined by interpolation of the LES data. Once the particles are initialized outside the solid boundary, no additional wall treatment is necessary to avoid the particles being advected to inside the airfoil since the flowfield already satisfies the no-slip and no-penetration condition. The interpolation is achieved by creating a piecewise cubic Bezier polynomial, utilizing the Clough-Tocher scheme \cite{ALFELD1984169, FARIN198683}.

\subsection{Compressibility effects in stability analysis}\label{app:B}

\begin{figure}[H]
    \centering
    \begin{overpic}[trim = 0cm 0cm 0cm 0cm,clip,width=0.79\textwidth]{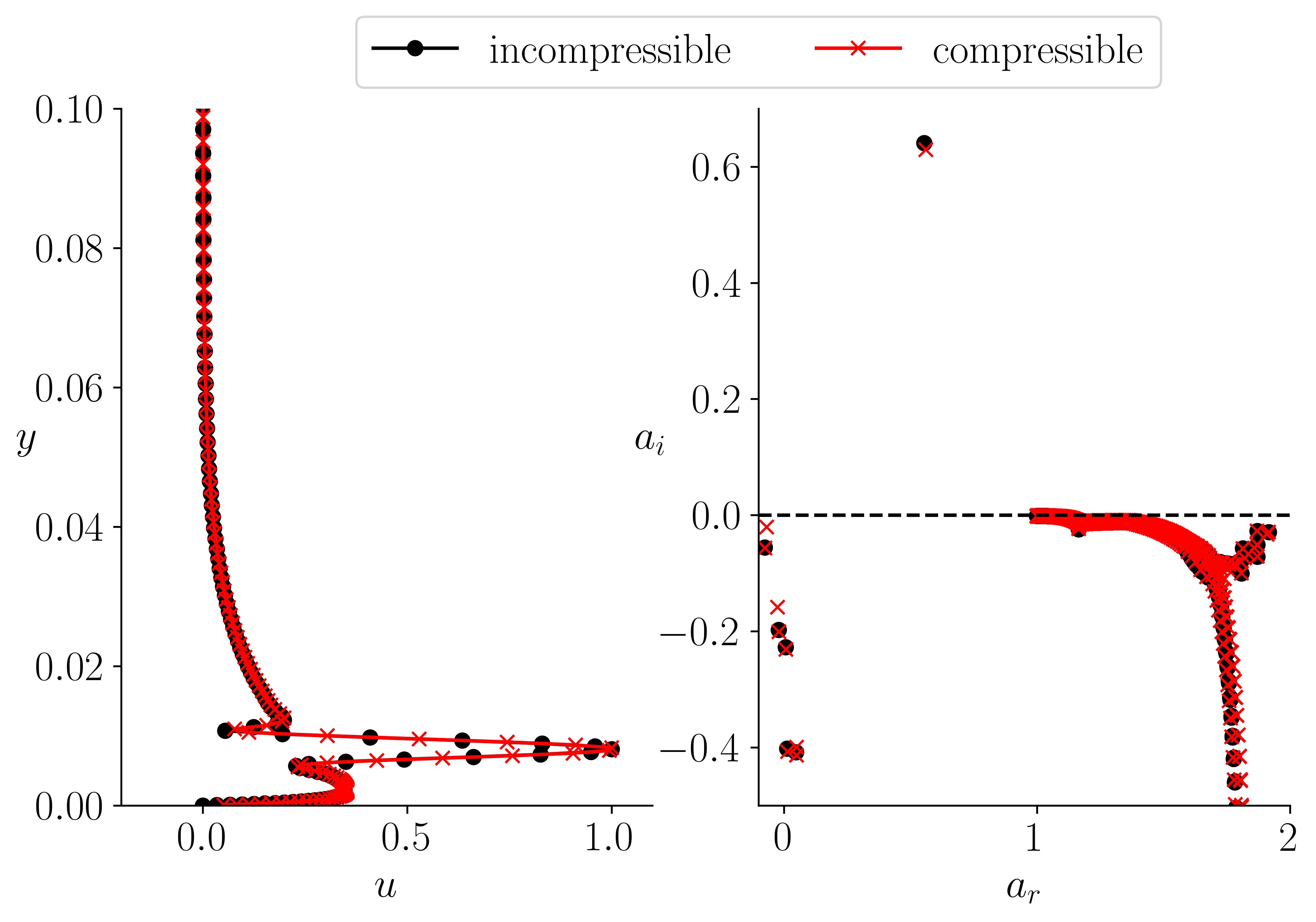}
    \end{overpic}
\caption{Unstable mode and spectra for compressible (red) and incompressible (black) linear operators.}
\label{fig : validation}
\end{figure}

In the present dynamic stall flow, the local Mach number may increase considerably due to the flow acceleration near the airfoil leading edge, as shown in Ref. \cite{Miotto_Wolf_Gaitonde_Visbal_2022}. Hence, in order to verify if these compressibility effects are important for the current stability analysis, we perform a comparison between results produced by compressible and a incompressible linear operators. The linearized compressible Navier-Stokes operator is formed using the finite-volume compressible flow solver \textit{CharLES} \cite{sun2017biglobal, rolandi2024invitation}. Boundary
conditions similar to those used in the incompressible linear operator are employed. The results are compared for an effective attack angle of $22^\circ$, as shown in Fig. \ref{fig : profiles 1}. The most unstable modes obtained from both operators are shown in Fig. \ref{fig : validation}, along with a comparison of the eigenvalue spectra. 
For the latter, one can see that the compressible operator spectrum, marked by the red $\times$ symbols, presents additional eigenvalues due to the energy equation that is also solved.
Overall, the results show good agreement with respect to both the spatial support of the mode and the eigenspectrum. Therefore, it is clear that, for the present flow, compressibility effects do not show a significant impact in the stability analysis. 

\section*{Funding Sources}

The authors acknowledge Fundação de Amparo à Pesquisa do Estado de São Paulo, FAPESP, for supporting the present work under research grants No. 2013/08293-7, 2019/17874-0, 2021/06448-0, 2022/08567-9, and 2023/13599-0. Conselho Nacional de Desenvolvimento Científico e Tecnológico, CNPq, is also acknowledged for supporting this research under grant No.\ 308017/2021-8.

\section*{Acknowledgments}

CEPID-CCES and CENAPAD-SP are acknowledged for providing the computational resources for this research through clusters Coaraci and Lovelace, respectively.

\bibliography{sample}

\end{document}